\pdfoutput=1

\documentclass[12pt,a4paper]{article}

\usepackage{ifthen} 
\newboolean{pdflatex}
\setboolean{pdflatex}{true} 

\newboolean{articletitles}
\setboolean{articletitles}{true} 

\newboolean{uprightparticles}
\setboolean{uprightparticles}{false} 

\usepackage{lscape}

\def\paperauthors{LHCb collaboration} 
\def\paperasciititle{Measurement of differential bb and cc dijet cross-sections in the forward region of pp collisions at 13 TeV} 
\def\papertitle{Measurement of differential \\ $b\bar{b}$- and $c\bar{c}$-dijet cross-sections \\ in the forward region of $pp$ collisions at $\sqrt{s}=13 ~ \mathrm{TeV}$} 
\def\paperkeywords{{High Energy Physics}, {LHCb}} 
\def\papercopyright{\the\year\ CERN for the benefit of the LHCb collaboration} 
\def\paperlicence{CC BY 4.0 licence}
\def\paperlicenceurl{https://creativecommons.org/licenses/by/4.0/}

\usepackage[top=1in, bottom=1.25in, left=1in, right=1in]{geometry}

\usepackage{multirow}
\usepackage{comment}
\usepackage{placeins}
\usepackage{siunitx,fixmath}

\usepackage{microtype}
\usepackage{lineno}  
\usepackage{xspace} 
\usepackage{caption} 

\usepackage{graphicx}  
\usepackage{color}
\usepackage{colortbl}
\graphicspath{{./figs/}} 
\DeclareGraphicsExtensions{.pdf,.PDF,png,.PNG}

\usepackage{amsmath} 
\usepackage{amssymb}
\usepackage{amsfonts}
\usepackage{upgreek} 
\usepackage{booktabs}

\newcommand*\patchAmsMathEnvironmentForLineno[1]{%
\expandafter\let\csname old#1\expandafter\endcsname\csname #1\endcsname
\expandafter\let\csname oldend#1\expandafter\endcsname\csname
end#1\endcsname
 \renewenvironment{#1}%
   {\linenomath\csname old#1\endcsname}%
   {\csname oldend#1\endcsname\endlinenomath}%
}
\newcommand*\patchBothAmsMathEnvironmentsForLineno[1]{%
  \patchAmsMathEnvironmentForLineno{#1}%
  \patchAmsMathEnvironmentForLineno{#1*}%
}
\AtBeginDocument{%
\patchBothAmsMathEnvironmentsForLineno{equation}%
\patchBothAmsMathEnvironmentsForLineno{align}%
\patchBothAmsMathEnvironmentsForLineno{flalign}%
\patchBothAmsMathEnvironmentsForLineno{alignat}%
\patchBothAmsMathEnvironmentsForLineno{gather}%
\patchBothAmsMathEnvironmentsForLineno{multline}%
\patchBothAmsMathEnvironmentsForLineno{eqnarray}%
}

\usepackage{hyperxmp}

\usepackage[pdftex,
            pdfauthor={\paperauthors},
            pdftitle={\paperasciititle},
            pdfkeywords={\paperkeywords},
            pdfcopyright={Copyright (C) \papercopyright},
            pdflicenseurl={\paperlicenceurl}]{hyperref}

\usepackage[colorinlistoftodos,textsize=scriptsize]{todonotes}

\usepackage[all]{hypcap} 

\usepackage{xspace} 
\usepackage{upgreek}

\def\lhcb   {\mbox{LHCb}\xspace}
\def\atlas  {\mbox{ATLAS}\xspace}
\def\cms    {\mbox{CMS}\xspace}

\def\MagUp {\mbox{\em Mag\kern -0.05em Up}\xspace}

\ifthenelse{\boolean{uprightparticles}}%
{

 \def\Pmu         {\ensuremath{\upmu}\xspace}

 \def\PDelta      {\ensuremath{\Delta}\xspace}                 
 \def\PXi         {\ensuremath{\Xi}\xspace}                 
 \def\PLambda     {\ensuremath{\Lambda}\xspace}                 
 \def\PSigma      {\ensuremath{\Sigma}\xspace}                 
 \def\POmega      {\ensuremath{\Omega}\xspace}                 
 \def\PUpsilon    {\ensuremath{\Upsilon}\xspace}

 \def\PB      {\ensuremath{\mathrm{B}}\xspace}                 
                  
 \def\PD      {\ensuremath{\mathrm{D}}\xspace}

 \def\PK      {\ensuremath{\mathrm{K}}\xspace}

 \def\PZ      {\ensuremath{\mathrm{Z}}\xspace}
                  
 \def\Pb      {\ensuremath{\mathrm{b}}\xspace}                 
 \def\Pc      {\ensuremath{\mathrm{c}}\xspace}

 \def\Pi      {\ensuremath{\mathrm{i}}\xspace}

 \def\Ps      {\ensuremath{\mathrm{s}}\xspace}

 \def\thebaroffset{0.0em}
}
{

 \def\Pmu         {\ensuremath{\mu}\xspace}

 \mathchardef\PDelta="7101
 \mathchardef\PXi="7104
 \mathchardef\PLambda="7103
 \mathchardef\PSigma="7106
 \mathchardef\POmega="710A
 \mathchardef\PUpsilon="7107
                  
 \def\PB      {\ensuremath{B}\xspace}                 
                  
 \def\PD      {\ensuremath{D}\xspace}

 \def\PK      {\ensuremath{K}\xspace}

 \def\PZ      {\ensuremath{Z}\xspace}                 
                  
 \def\Pb      {\ensuremath{b}\xspace}                 
 \def\Pc      {\ensuremath{c}\xspace}

 \def\Pi      {\ensuremath{i}\xspace}

 \def\Ps      {\ensuremath{s}\xspace}

 \def\thebaroffset{0.18em}
}
\newcommand{\offsetoverline}[2][\thebaroffset]{\kern #1\overline{\kern -#1 #2}}%
\newcommand\Zpjet   {\ensuremath{Z +\mathrm{jet}}\xspace}

\makeatletter
\ifcase \@ptsize \relax
  \newcommand{\miniscule}{\@setfontsize\miniscule{4}{5}}
\or
  \newcommand{\miniscule}{\@setfontsize\miniscule{5}{6}}
\or
  \newcommand{\miniscule}{\@setfontsize\miniscule{5}{6}}
\fi
\makeatother

\DeclareRobustCommand{\optbar}[1]{\shortstack{{\miniscule (\rule[.5ex]{1.25em}{.18mm})}
  \\ [-.7ex] $#1$}}

\def\mup        {{\ensuremath{\Pmu^+}}\xspace}
\def\mun        {{\ensuremath{\Pmu^-}}\xspace} 

\def\squark    {{\ensuremath{\Ps}}\xspace}

\def\cquark    {{\ensuremath{\Pc}}\xspace}
\def\cquarkbar {{\ensuremath{\overline \cquark}}\xspace}
\def\ccbar     {{\ensuremath{\cquark\cquarkbar}}\xspace}
\def\bquark    {{\ensuremath{\Pb}}\xspace}
\def\bquarkbar {{\ensuremath{\overline \bquark}}\xspace}
\def\bbbar     {{\ensuremath{\bquark\bquarkbar}}\xspace}

\def\KorKbar {\kern \thebaroffset\optbar{\kern -\thebaroffset \PK}{}\xspace}

\def\DorDbar {\kern \thebaroffset\optbar{\kern -\thebaroffset \PD}\xspace}

\def\B       {{\ensuremath{\PB}}\xspace}

\def\BorBbar {\kern \thebaroffset\optbar{\kern -\thebaroffset \PB}\xspace}

\def\Bd      {{\ensuremath{\B^0}}\xspace}

\def\BdorBdbar {\kern \thebaroffset\optbar{\kern -\thebaroffset \Bd}\xspace}

\def\Bs      {{\ensuremath{\B^0_\squark}}\xspace}

\def\BsorBsbar {\kern \thebaroffset\optbar{\kern -\thebaroffset \Bs}\xspace}

\def\Y#1S{\ensuremath{\PUpsilon{(#1S)}}\xspace}

\def\LorLbar     {\kern \thebaroffset\optbar{\kern -\thebaroffset \PLambda}\xspace}

\def\AT#1     {\ensuremath{A_{\mathrm{T}}^{#1}}\xspace}           

\def\C#1      {\ensuremath{\mathcal{C}_{#1}}\xspace}                       
\def\Cp#1     {\ensuremath{\mathcal{C}_{#1}^{'}}\xspace}                    
\def\Ceff#1   {\ensuremath{\mathcal{C}_{#1}^{\mathrm{(eff)}}}\xspace}        
\def\Cpeff#1  {\ensuremath{\mathcal{C}_{#1}^{'\mathrm{(eff)}}}\xspace}       
\def\Ope#1    {\ensuremath{\mathcal{O}_{#1}}\xspace}                       
\def\Opep#1   {\ensuremath{\mathcal{O}_{#1}^{'}}\xspace}                    

\newcommand{\nospaceunit}[1]{\ensuremath{\text{#1}}}       
\newcommand{\aunit}[1]{\ensuremath{\text{\,#1}}}       

\newcommand{\tev}{\aunit{Te\kern -0.1em V}\xspace}
\newcommand{\gev}{\aunit{Ge\kern -0.1em V}\xspace}
\newcommand{\mev}{\aunit{Me\kern -0.1em V}\xspace}
\newcommand{\kev}{\aunit{ke\kern -0.1em V}\xspace}
\newcommand{\ev}{\aunit{e\kern -0.1em V}\xspace}
\newcommand{\mevc}{\ensuremath{\aunit{Me\kern -0.1em V\!/}c}\xspace}
\newcommand{\gevc}{\ensuremath{\aunit{Ge\kern -0.1em V\!/}c}\xspace}
\newcommand{\mevcc}{\ensuremath{\aunit{Me\kern -0.1em V\!/}c^2}\xspace}
\newcommand{\gevcc}{\ensuremath{\aunit{Ge\kern -0.1em V\!/}c^2}\xspace}

\def\mum  {\ensuremath{\,\upmu\nospaceunit{m}}\xspace}

\def\nb {\aunit{nb}\xspace}

\def\fb   {\ensuremath{\aunit{fb}}\xspace}
\def\invfb   {\ensuremath{\fb^{-1}}\xspace}

\def\deriv {\ensuremath{\mathrm{d}}}

\def\gsim{{~\raise.15em\hbox{$>$}\kern-.85em
          \lower.35em\hbox{$\sim$}~}\xspace}
\def\lsim{{~\raise.15em\hbox{$<$}\kern-.85em
          \lower.35em\hbox{$\sim$}~}\xspace}

\def\pt         {\ensuremath{p_{\mathrm{T}}}\xspace}

\def\ptot       {\ensuremath{p}\xspace}

\def\evtgen     {\mbox{\textsc{EvtGen}}\xspace}

\def\geant      {\mbox{\textsc{Geant4}}\xspace}

\def\photos     {\mbox{\textsc{Photos}}\xspace}

\def\pythia     {\mbox{\textsc{Pythia}}\xspace}

\def\fastjet       {\mbox{\textsc{Fastjet}}\xspace}

\xspace

\def\tell1  {TELL1\xspace}
\def\ukl1   {UKL1\xspace}

\usepackage{cite} 
\usepackage{mciteplus}

\usepackage{longtable} 
\usepackage{appendix}

\begin{document}

\renewcommand{\thefootnote}{\fnsymbol{footnote}}
\setcounter{footnote}{1}


\begin{titlepage}
\pagenumbering{roman}

\vspace*{-1.5cm}
\centerline{\large EUROPEAN ORGANIZATION FOR NUCLEAR RESEARCH (CERN)}
\vspace*{1.5cm}
\noindent
\begin{tabular*}{\linewidth}{lc@{\extracolsep{\fill}}r@{\extracolsep{0pt}}}
\ifthenelse{\boolean{pdflatex}}
{\vspace*{-1.5cm}\mbox{\!\!\!\includegraphics[width=.14\textwidth]{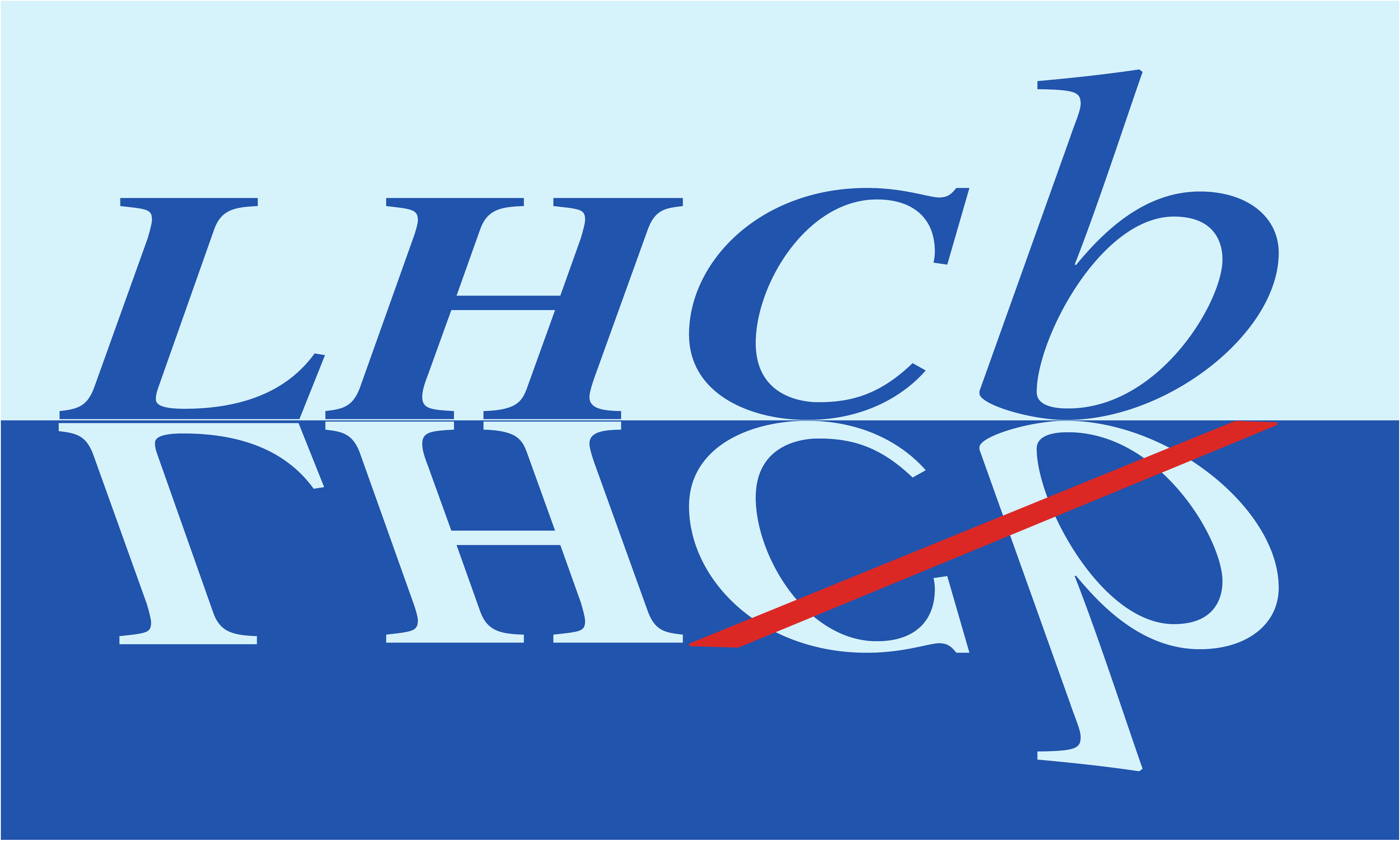}} & &}%
{\vspace*{-1.2cm}\mbox{\!\!\!\includegraphics[width=.12\textwidth]{lhcb-logo.eps}} & &}%
\\
 & & CERN-EP-2020-174 \\  
 & & LHCb-PAPER-2020-018 \\  
 & & 19 October 2020 \\ 
 & & \\
\end{tabular*}

\vspace*{2.0cm}

{\normalfont\bfseries\boldmath\huge
\begin{center}
  \papertitle 
\end{center}
}

\vspace*{1.5cm}

\begin{center}
\paperauthors\footnote{Authors are listed at the end of this paper.}
\end{center}

\vspace{\fill}

\begin{abstract}
  \noindent
  The inclusive $b \bar{b}$- and $c \bar{c}$-dijet production cross-sections in the forward region of $pp$ collisions are measured using a data sample collected with the LHCb detector at a centre-of-mass energy of 13\tev in 2016. The data sample  corresponds to an integrated luminosity of 1.6\invfb.
  Differential cross-sections are measured as a function of the transverse momentum and of the pseudorapidity of the leading jet,  of the rapidity difference between the jets, and  of the dijet invariant mass.
  A fiducial region for the measurement is defined by requiring that the two jets originating from the two $b$ or $c$ quarks are emitted with transverse momentum greater than 20\gevc, pseudorapidity in the range $2.2 < \eta < 4.2$, and with a difference in the azimuthal angle between the two jets greater than 1.5. The integrated $b \bar{b}$-dijet cross-section is measured to be $53.0 \pm 9.7~\nb$, and the total $c \bar{c}$-dijet cross-section is measured to be $73 \pm 16~\nb$. The ratio between $c \bar{c}$- and $b \bar{b}$-dijet cross-sections is also measured and  found to be $1.37 \pm 0.27$.
  The results are in agreement with  theoretical predictions at next-to-leading order.
  
\end{abstract}

\vspace*{2.0cm}

\begin{center}
  Published in JHEP 02 (2021) 023 
\end{center}

\vspace{\fill}

{\footnotesize 
\centerline{\copyright~\papercopyright. \href{\paperlicenceurl}{\paperlicence}.}}
\vspace*{2mm}

\end{titlepage}


\newpage
\setcounter{page}{2}
\mbox{~}
%
%
%
%

\cleardoublepage


\renewcommand{\thefootnote}{\arabic{footnote}}
\setcounter{footnote}{0}


\pagestyle{plain} 
\setcounter{page}{1}
\pagenumbering{arabic}


\clearpage


\section{Introduction}
\label{sec:Introduction}
Measurements of  $b\bar{b}$ and $c\bar{c}$ production cross-sections provide an important test of quantum chromodynamics (QCD) in proton-proton collisions.
In these collisions, bottom and charm quarks are mostly produced in pairs by quark and gluon scattering processes, predominantly by flavour creation, flavour excitation and gluon splitting~\cite{Cacciari:2012ny,*Kniehl:2011bk}. 
At the LHC energies beauty and charm quarks produced in the collisions are likely to generate jets through fragmentation and hadronization processes.
Experimentally, one can infer the production of a beauty (charm) quark either through exclusively identifying $b$ ($c$) hadron decays, or through the reconstruction of jets that are tagged as originating in heavy flavour quark fragmentation. 

As $b\bar{b}$- and $c\bar{c}$-dijet differential cross-sections can be calculated in perturbative QCD (pQCD) as a function of the dijet kinematics, comparisons between data and predictions provide a critical test of next-to-leading-order (NLO) pQCD calculations~\cite{madgraph,hadroproduction}. 
Measurements of differential cross-sections of heavy-flavour dijets can also be a sensitive probe of the parton distribution functions (PDFs) of the proton. 
Among the LHC experiments, the PDF region with low Bjorken-$x$ values is accessible only to LHCb, due to its forward acceptance~\cite{Rojo:2017xpe}.
Moreover, the knowledge of the inclusive $b$ and $c$ quarks production rate from QCD processes is necessary to understand the background contributions in searches for massive particles decaying into $b$ or $c$ quarks, such as the Higgs boson or new heavy particles.

In 2013, the LHCb collaboration measured the integrated $b\bar{b}$ and $c\bar{c}$ production cross-sections at a centre-of-mass energy of \mbox{$\sqrt{s}=$ 7~\tev} in the region of pseudorapidity $2.5 < \eta < 4.0$, tagging the quark flavour via the reconstruction of displaced vertices \cite{LHCB-CONF-2013-002}. The LHCb collaboration has also measured the $b$-quark production cross-section at $\sqrt{s}=7$ and 13~\tev in the region with pseudorapidity $2 < \eta < 5$, using semileptonic decays of $b$-flavoured hadrons~\cite{LHCb-PAPER-2016-031}.
The \atlas collaboration has measured the  inclusive $b\bar{b}$-dijet production cross-section~\cite{Aaboud:2016jed}, while the \cms collaboration has performed a measurement of the inclusive $b$-jet production cross-section~\cite{Chatrchyan:2012dk}. The latter two measurements were performed at \mbox{$\sqrt{s}=7$ \tev} in the central pseudorapidity region, with $|\eta| < 2.5$.

In this paper, a measurement of the inclusive $b\bar{b}$- and $c\bar{c}$-dijet cross-sections at \mbox{$\sqrt{s}=13$ \tev} is presented. The data sample used corresponds to a total integrated luminosity of proton--proton ($pp$) collisions of 1.6 $\mathrm{fb}^{-1}$, collected during the year 2016.  Cross-section measurements are also performed differentially as a function of the dijet kinematics. The ratio of the $c \bar{c}$ to the $b \bar{b}$ cross-sections is also determined.
This is the first $c\bar{c}$-dijet differential cross-section measurement at a hadron collider.
 
This paper is structured as follows. The \lhcb detector and the  simulation samples used in this analysis are introduced in Sec.~\ref{sec:Detector}. Section~\ref{sec:selection} presents the selection of  the  events and the tagging of jets as originating from $b$ and $c$ quarks, as well as the definition  of the variables  used for the cross-section  measurement.  The fitting procedure is described  in Sec.~\ref{sec:fit}. The unfolding procedure used to convert the raw observables into generator-level observables is described in  Sec.~\ref{sec:unfolding}.  Systematic uncertainties on the cross-section measurements are discussed in Sec.~\ref{sec:systematics}. The determination of the cross-section ratios is introduced in Sec.~\ref{sec:ratios}. Finally, results are shown in Sec.~\ref{sec:results} and conclusions are drawn in Sec.~\ref{sec:Conclusions}.

\section{Detector and simulation}
\label{sec:Detector}

The \lhcb detector~\cite{Alves:2008zz,LHCb-DP-2014-002} is a single-arm forward
spectrometer covering the \mbox{pseudorapidity} range $2<\eta <5$,
designed for the study of particles containing \bquark or \cquark
quarks. The detector includes a high-precision tracking system
consisting of a silicon-strip vertex detector surrounding the $pp$
interaction region~\cite{LHCb-DP-2014-001}, a large-area silicon-strip detector located
upstream of a dipole magnet with a bending power of about
$4{\mathrm{\,Tm}}$, and three stations of silicon-strip detectors and straw
drift tubes~\cite{LHCb-DP-2017-001} placed downstream of the magnet.
The tracking system provides a measurement of the momentum, \ptot, of charged particles with
a relative uncertainty that varies from 0.5\% at low momentum to 1.0\% at 200\gevc.
The minimum distance of a track to a primary vertex (PV), the impact parameter (IP), 
is measured with a resolution of $(15+29/\pt)\mum$,
where \pt is the component of the momentum transverse to the beam, in\,\gevc.
Different types of charged hadrons are distinguished using information
from two ring-imaging Cherenkov detectors~\cite{LHCb-DP-2012-003}. 
Photons, electrons and hadrons are identified by a calorimeter system consisting of
scintillating-pad and preshower detectors, an electromagnetic
and a hadronic calorimeter. Muons are identified by a
system composed of alternating layers of iron and multiwire
proportional chambers~\cite{LHCb-DP-2012-002}. The online event selection is performed by a trigger~\cite{LHCb-DP-2012-004}, 
which consists of a hardware stage, based on information from the calorimeter and muon
systems, followed by a software stage, which applies a full event
reconstruction.

 At the hardware trigger stage, events for this analysis are required to contain a reconstructed muon with high \pt or a
 hadron, photon or electron with high transverse energy in the calorimeters. 
 A global event cut (GEC) on the number of hits in the scintillating-pad detector is also applied.
 The software trigger requires at least one charged particle
 to be reconstructed with $\pt > 1.6\gevc$ that is 
 inconsistent with originating from any PV, as well as the presence of two jets.  
 Both jets are reconstructed as described below, and required to have $\pt > 17\gevc$ and a secondary vertex (SV) in the jet cone.

 Simulation is required to model and correct for the effects of the detector acceptance and the
  imposed selection requirements.
  In the simulation, $pp$ collisions are generated using
\pythia~\cite{Sjostrand:2007gs,*Sjostrand:2006za} 
 with a specific \lhcb configuration~\cite{LHCb-PROC-2010-056}.  Decays of unstable particles
are described by \evtgen~\cite{Lange:2001uf}, in which final-state
radiation is generated using \photos~\cite{Golonka:2005pn}. The
interaction of the generated particles with the detector, and its response,
are implemented using the \geant
toolkit~\cite{Allison:2006ve, *Agostinelli:2002hh} as described in
Ref.~\cite{LHCb-PROC-2011-006}.

This work uses simulated samples of $b \bar{b}$-dijets, $c \bar{c}$-dijets and dijets generated from light partons ($u$, $d$, $s$ quarks, and gluons, indicated in the following by $q$). These samples are used 
to model the distributions of observables employed in the heavy flavour identification, to measure the $b$- and $c$-jet selection efficiencies and to  determine the unfolding matrices used to convert to generator-level quantities.
In order to cover the full range of jet \pt, several simulated samples with different values of transverse momentum exchanged in the hard interaction ($\hat{p}_{\mathrm{T}}$) are generated. When combining the samples appropriate weights are used, depending on the range of $\hat{p}_{\mathrm{T}}$. These weights are taken to be proportional to the cross-sections evaluated with \pythia \cite{Sjostrand:2007gs,*Sjostrand:2006za} for the different $\hat{p}_{\mathrm{T}}$ ranges.


\section{Jet reconstruction and event selection}
\label{sec:selection}

Jets are reconstructed using particle flow objects as input \cite{LHCb-PAPER-2013-058}.
The objects are combined employing the anti-$k_{\rm T}$ algorithm \cite{antikt}, as implemented in the \fastjet software package \cite{fastjet}, with a jet radius parameter of $R=0.5$. 
The offline and online jet reconstruction algorithms are identical, however minor differences between offline and online may arise from different reconstruction routines for tracks and calorimeter clusters that are used in the two contexts. Systematic uncertainties are evaluated to cover these small differences and described in Sec.~\ref{sec:systematics}. 

To improve the rejection of fake jets, such as jets originating from noise and high energy isolated leptons, additional criteria, similar to those explained in Ref.~\cite{LHCb-PAPER-2013-058}, are imposed.
In particular jets are required to contain at least two particles matched to the same PV, at least one track with $p_{\mathrm{T}}>1.2$ GeV, no single particle with more than 10\% of the jet $p_{\mathrm{T}}$ and to have the fraction of the jet $p_{\mathrm{T}}$ carried by charged particles greater than 10\%.
These requirements have been optimized using simulated samples produced with 2016 running conditions.

 In this paper the jet flavours are distinguished by using a heavy-flavour jet-tagging algorithm, that is referred to as ``SV-tagging''. The SV-tagging algorithm reconstructs secondary vertices (SVs) using tracks inside and outside of the jet and is described in detail in Ref.~\cite{LHCb-PAPER-2015-016}. In this algorithm tracks that have a significant \pt and displacement from every PV are combined to form two-body SVs. Then good quality two-body SVs are linked together if they share one track, in order to form $n$-body SVs.
If a SV is found inside the cone of the jet, the jet is tagged as likely to be originating from $b$- or $c$-quark fragmentation.
The SV-tagging efficiency, determined in simulation, is about 60\% for $b$-jets and 20\% for $c$-jets. These values are lower with respect to those obtained in the $\sqrt{s}=7$ and 8 \tev datasets studied in Ref.~\cite{LHCb-PAPER-2015-016}. The relative efficiency loss is below the 10\% level, and is explained by the higher particle multiplicity of $\sqrt{s}=13$ \tev events with respect to $\sqrt{s}=7$ and 8 \tev events, that introduces more noise in the SV finding.
To further distinguish light-flavour jets from heavy-flavour jets and $b$-jets from $c$-jets multivariate analysis algorithms as described in Ref.~\cite{LHCb-PAPER-2015-016} are used.
Two boosted decision tree (BDT) classifiers \cite{Breiman, Roe, AdaBoost}, that use as inputs variables related to the SV, are employed: one for heavy-/light-jet separation (BDT$_{bc|q}$) and the other for  $b$-/$c$-jet separation (BDT$_{b|c}$).

The offline selection is applied to events that pass the trigger criteria for heavy-flavour dijets. 
Two offline-reconstructed jets originating from the same PV are selected as dijet candidates.
The kinematic requirements in Tab.~\ref{tab:fiducial_region} are applied to the reconstructed jets. In the table and in the remainder of the paper, the leading jet, $j_0$, is that with the largest $p_{\mathrm{T}}$, and $j_1$ is the other jet in the pair.  The kinematic selection includes a requirement on  $|\Delta \phi|$,  the difference in the azimuthal angle between the jets.
In 0.4$\%$ of the selected events multiple dijet candidates exist after applying all the requirements; the jet pair with maximum sum of the $p_{\mathrm{T}}$ of the two jets is selected in these cases. It has been verified in simulation that this choice does not bias the results.
The fraction of events with multiple candidates found in simulation is similar to that in data.
The differential cross-section is measured as a function of four observables: the leading jet pseudorapidity $\eta(j_0)$, the leading jet transverse momentum $\pt(j_0)$, the dijet invariant mass,  $m_{jj}$, and 
\begin{displaymath}
\Delta y^{*} = \frac{1}{2}|y_0 - y_1|,
\end{displaymath}
where $y_0$ and $y_1$ are the jet rapidities.

\begin{table}[t]
  \begin{center}
      \caption{\label{tab:fiducial_region} List of fiducial requirements on jet transverse momentum, pseudorapidity and the azimuthal angle between the jets.}
      \begin{tabular}{c}
        \toprule
        $p_{\mathrm{T}}(j_0)>20$ \gevc\\
        $p_{\mathrm{T}}(j_1)>20$ \gevc\\
        $2.2 < \eta(j_0) < 4.2$\\
        $2.2 < \eta(j_1) < 4.2$\\
        $|\Delta \phi|>1.5$ \\ 
        \bottomrule
      \end{tabular}
  \end{center}
\end{table}

Finally, a data sample in which a \PZ boson is  produced in association with a jet and decays to a $\mup \mun$ pair is used to measure efficiencies and assess several systematic uncertainties in this analysis.
A similar selection to that of Ref.~\cite{LHCB-PAPER-2016-011} is applied, with some differences introduced to match the jet  phase space considered in this analysis. This sample is further referred to as \Zpjet.


\section{Fitting procedure}
\label{sec:fit}

A fit to SV-tagging-related observables is performed in order to extract the $b \bar{b}$- and $c \bar{c}$-dijet yields. The fit is performed in intervals of the dijet kinematics introduced in the previous section. 
The expected distributions of the tagging observables for $b \bar{b}$, $c \bar{c}$ and background samples are obtained as histograms using simulated samples.
Four tagging observables are used for disentangling the $b\bar{b}$ and $c \bar{c}$ processes from the background: the output of the classifiers $\text{BDT}_{bc|q}$ and $\text{BDT}_{b|c}$ for the leading jet and $\text{BDT}_{bc|q}$ and $\text{BDT}_{b|c}$ for the second jet. In principle a four-dimensional fit would give the best result in terms of the statistical uncertainty, but this is not optimal given the finite simulated sample sizes. Instead, two new observables are built, introducing linear combinations of the four tagging observables:
\begin{equation}
    \begin{aligned}
        t_0 &= \text{BDT}_{bc|q}(j_0) + \text{BDT}_{bc|q}(j_1),\\
        t_1 &= \text{BDT}_{b|c}(j_0) + \text{BDT}_{b|c}(j_1).
    \end{aligned}
    \end{equation}
    An alternative method, where the multiplication of SV-tagging observables is considered instead of the sum, is used to evaluate a systematic uncertainty on the procedure.

\begin{figure}[tbp]
  \begin{center}
    \includegraphics[width=0.48\linewidth]{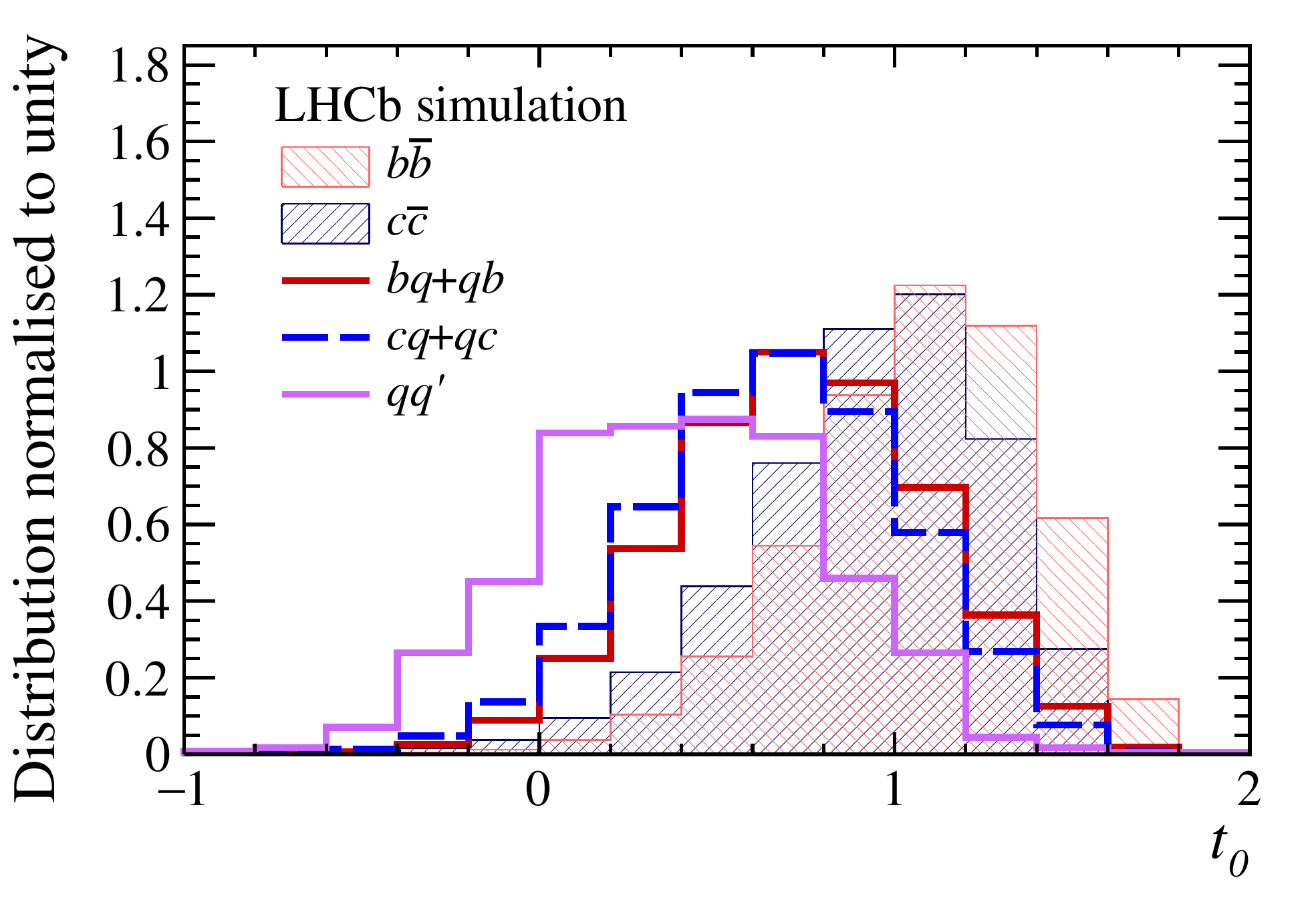} \hspace{.25cm}
    \includegraphics[width=0.48\linewidth]{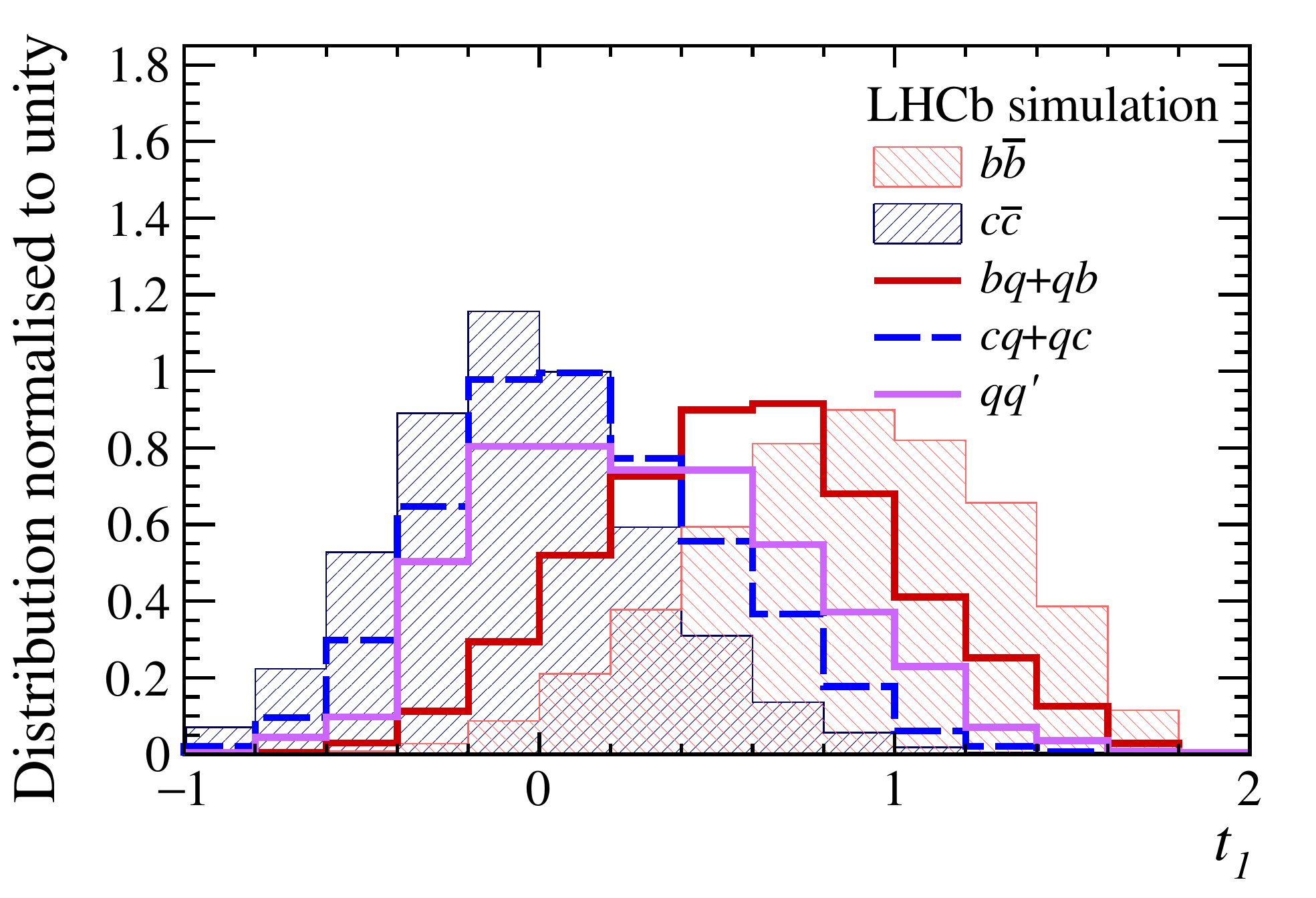}
  \end{center}
  \caption{
    Fit templates for (left) $t_0$ and (right) $t_1$ projections for the same-flavour and different-flavour processes, summed over the $[\pt(j_0),\pt(j_1)]$ bins, where $t_0$ and $t_1$ are the linear  combinations of the  tagging observables of both jets. The templates are normalised to unit area. To simplify the visualisation, the $qb$ and $bq$ samples ($qc$ and $cq$) are merged in the plot.
    }
  \label{fig:x0x1}
\end{figure}

Three different types of processes are expected in the data sample: same-flavour processes, different-flavour processes and background from light jets.
In same-flavour processes two $b$-jets or two $c$-jets are detected in the acceptance, they are labeled as $b\bar b$ and $c\bar c$. 
Different-flavour processes are $b\bar{b} q$ and $c\bar{c} q$ processes (with $q=u,d,s,g$) where one $b$- or $c$-jet is detected in the \lhcb acceptance and the second jet in the dijet is a light flavour jet, they are labeled as $bq$ and $cq$.  The $b\bar{b}c\bar{c}$ process has a cross-section of about three orders of magnitude smaller than the $b\bar{b} q$ and $c\bar{c} q$ processes \cite{Sjostrand:2007gs,*Sjostrand:2006za}, and is neglected in the fit. 
The background from light jets, where the two light jets may have different flavours, is labeled as $qq'$.  
Fit templates are constructed as two-dimensional $(t_0,t_1)$ histograms, with a $20 \times 20$ binning scheme and $t_0, t_1 \in [-2,2]$. For the same-flavour processes and the background from light jets, the histograms are filled with simulated events. 
For the different-flavour processes, two-dimensional ($\text{BDT}_{bc|q}$,$\text{BDT}_{b|c}$) single-jet templates with $20 \times 20$ bins are built using the $b\bar{b}$, $c\bar{c}$ and light partons simulation samples. Two-dimensional $(t_0,t_1)$ different-flavour templates are then obtained from a  convolution of two single-jet templates. 
Same-flavour, different-flavour and light jets template projections in $t_0$ and $t_1$ obtained in this way are shown in Fig.~\ref{fig:x0x1}. For different-flavour processes, separate $bq$, $qb$, $cq$, $qc$ templates are considered where the first flavour is associated to $j_0$ and the second to $j_1$.

The outputs of the $\text{BDT}_{bc|q}$ and $\text{BDT}_{b|c}$ classifiers show a correlation with the jet \pt, while they are almost uncorrelated with the jet pseudorapidity and the other kinematic variables.
For this reason different templates are built for different intervals of $[\pt(j_0),\pt(j_1)]$. 
The \pt binning scheme is the following: [20,30]\gevc, [30,40]\gevc, [40,50]\gevc, [50,60]\gevc and $\pt>60$ \gevc. As by definition $\pt(j_0)$ is higher than $\pt(j_1)$, only 15 non-empty $[\pt(j_0),\pt(j_1)]$ intervals are present. 

In order to measure the differential $b\bar{b}$- and $c\bar{c}$-dijet yields the dataset is divided in sub-samples for each of the kinematic observables. For a given observable the data sample is divided in three dimensions into bins of that observable and in bins of $[\pt(j_0),\pt(j_1)]$. The $[\pt(j_0),\pt(j_1)]$ binning scheme is identical to that employed in the template construction. For each bin a $(t_0,t_1)$ fit is performed using templates corresponding to the $[\pt(j_0),\pt(j_1)]$ interval and the extracted $b\bar{b}$- and $c\bar{c}$-dijet yields are summed over the  $[\pt(j_0),\pt(j_1)]$ bins, in order to obtain the yields for the different kinematic observable intervals.
The fits are performed with the yield of each species as a freely varying parameter and each bin is fitted independently.  

\begin{figure}[htb]
  \begin{center}
    \includegraphics[width=1.\linewidth]{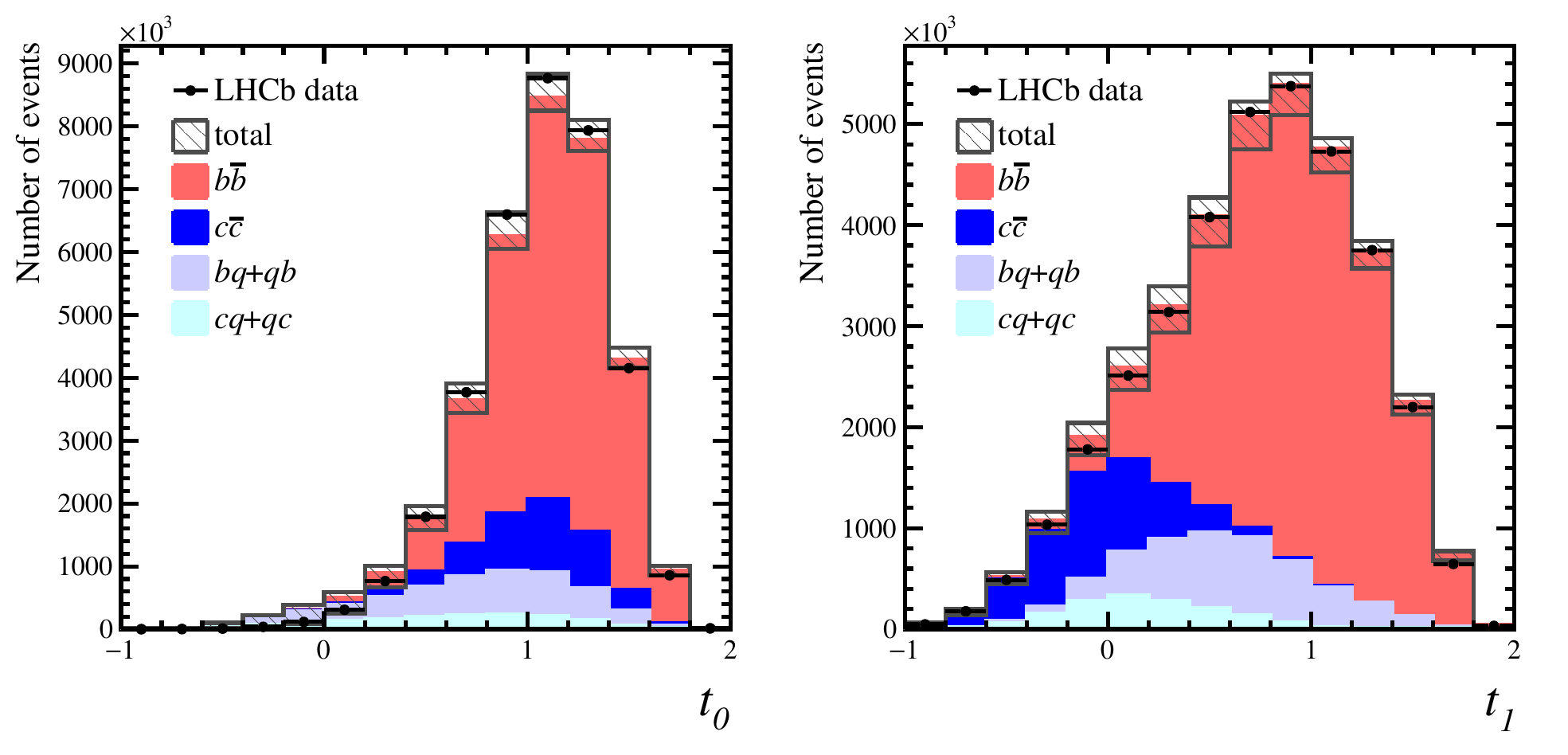}
  \end{center}
  \caption{
    The observables $t_0$ and $t_1$ obtained by summing the fitted yields over the $\eta(j_0)$ and  $[\pt(j_0),\pt(j_1)]$ bins, where $t_0$ and $t_1$ are the linear  combinations of the  tagging observables of both jets. The fit is compared against the data (black points). The statistical uncertainty on data is small and not visible in the plot. The stacked histograms show the contribution from: (red) $b\bar{b}$, (blue) $c\bar{c}$, (lavender) $bq$ and $qb$, (light blue) $cq$ and $qc$. The $qq'$ component is not displayed since its yield is negligible. The dashed grey areas represent the total uncertainty (statistical and systematic) on the fit result.}
  \label{fig:fit_result}
\end{figure}

The results obtained by summing the fitted yields of $b\bar{b}$, $c\bar{c}$, $bq+qb$ and $cq+qc$ over the $\eta(j_0)$ and $[\pt(j_0),\pt(j_1)]$ bins are shown in Fig.~\ref{fig:fit_result}. The uncertainty on the fit includes the statistical uncertainty on the data and systematic uncertainties related to the fit procedure, the template modeling and the finite size of the simulation samples used to construct the templates. The evaluation of these systematic uncertainties is described in Sec.~\ref{sec:systematics}.

Pseudoexperiments are performed to assess the fit stability and to determine the fit bias and coverage. Relative biases of the order of 0.01\% (0.02\%) on the fitted $b\bar{b}$ ($c \bar{c}$) yields are found, these values are used to correct the fit result. Moreover, the pseudoexperiments indicate that relative biases of the order $10\%$ on the fitted yield uncertainties are present. A correction is therefore also applied to the fit uncertainty.

\section{Determination of the cross-section}
\label{sec:unfolding}

The yields in each of the bins of the observables are used to calculate differential cross-sections at generator level, using an unfolding technique to correct for bin migrations due to detector effects and resolution. A least square method with Tikhonov regularisation \cite{Tikhonov:1963} is employed.
Generator-level jets are defined as jets clustered with the anti-$k_{\rm T}$ algorithm \cite{antikt} using generated quasi-stable particles, excluding neutrinos, as inputs.
The fiducial region of the measurement is defined by the kinematic requirements in Tab.~\ref{tab:fiducial_region} applied to jet observables at generator level.
The cross-sections are evaluated using 
\begin{equation}
\label{eq:cross}
\frac{\deriv\sigma}{\deriv z}(i) = \frac{1}{\mathcal{L}}  \cdot \frac{1}{2\Delta z(i)}  \cdot \frac{1}{A(i) \cdot \epsilon(i)}  \sum\limits_{j=1}^{n} U_{ij} \cdot N(j),
\end{equation}
where $z$ is the variable under study at generator level, $i$ indicates the index of the bin $[z_i - \Delta z(i), z_i + \Delta z(i)]$, $2\Delta z(i)$ is the width of the bin, $\mathcal{L}$ is the integrated luminosity, $N(j)$ is the number of fitted events in the bin $[z^{\text{reco}}_j - \Delta z^{\text{reco}}, z^{\text{reco}}_j + \Delta z^{\text{reco}}]$ defined using  the reconstructed variables, $A(i)$ is the acceptance factor for the bin $i$, $\epsilon(i)$ is the efficiency for the bin $i$ and $U_{ij}$ is the unfolding matrix that maps reconstructed to generator-level variables.
The acceptance factor is introduced in the cross-section formula to account for the migration of events in to and out of the fiducial region, which is not accounted for in the unfolding matrix.  

The efficiency is written as the product
\begin{displaymath}
\epsilon_{\text{tot}} = \epsilon_{\text{reco}} \cdot  \epsilon_{\text{tag}} \cdot \epsilon_{\text{trig}}, 
\end{displaymath}
where $\epsilon_{\text{reco}}$ is the jet reconstruction efficiency, $\epsilon_{\text{tag}}$ is the jet tagging efficiency of reconstructed jets and $\epsilon_{\text{trig}}$ is the trigger efficiency evaluated on tagged jets.

The total efficiency is obtained using simulated $b\bar{b}$ and $c\bar{c}$  samples. Per-event weights are applied to simulated events in order to correct for data/simulation  differences. For the trigger efficiency, $\epsilon_{\text{trig}}$, per-jet data/simulation  weights are measured following the procedure in Ref.~\cite{top}. The trigger efficiency must be also corrected for the GEC requirement, since its efficiency is about a factor 0.6 lower in data than in simulation.
The GEC efficiency is determined in data using independent samples of events with no  GECs  applied. Finally $\epsilon_{\text{trig}}$ is corrected for data/simulation  differences in the efficiency of the other trigger requirements. To do this per-jet weights are measured with a tag-and-probe technique, comparing data and simulation samples of \Zpjet events. 
The total selection efficiency for the $b\bar{b}$ process is found to be about $15\%$ and  about $1.5\%$ for the $c\bar{c}$ process. This difference is due to the SV-tagging efficiency, as explained in Sec. \ref{sec:selection}.

Unfolding matrices are obtained for each of the four considered observables using simulation.
Uncertainties due to the finite simulated sample size in the unfolding matrix construction are propagated to the result.
Since the detector response is known to be similar for $b$- and $c$-jets and the $b \bar{b}$ simulation sample is larger, the $b \bar{b}$ sample is used for both the $b \bar{b}$ and $c \bar{c}$ unfolding. A systematic uncertainty due to differences in the underlying dijet kinematic for $b \bar{b}$ and $c \bar{c}$ is discussed in Sec.~\ref{sec:systematics}.
%


\section{Systematic uncertainties}
\label{sec:systematics}

Systematic uncertainties can affect the fitting procedure, the selection efficiencies, the acceptance factor, the unfolding and the integrated luminosity.

The systematic uncertainty affecting the GEC efficiency arises mainly from the different values obtained in data subsamples where no GECs are applied, since these have different compositions of $b\bar{b}$, $c\bar{c}$ and $qq'$ events. The resulting uncertainty on the efficiency determination is 6.3\%, correlated across all bins. 
Remaining differences between data and simulation in the trigger efficiency are taken into account using per-event weights. The statistical uncertainty on the weights is taken as a systematic uncertainty. The mean relative uncertainty associated to these weights is around 3\%. An additional source of systematic uncertainty arises from the difference between the online and offline physical objects reconstruction algorithms. A subset of data events where only one jet is required to be SV-tagged at trigger level is used to assess this uncertainty. This systematic uncertainty is again around 3\%.

Data-simulation corrections are applied to simulated events in the evaluation of the SV-tagging efficiency.
The corrections and corresponding uncertainties follow Ref.~\cite{LHCb-PAPER-2015-016}, in which these values were computed with data taken at $\sqrt{s}= 7$ and 8\tev. 
A tag-and-probe technique was used on control samples with a jet and a $W$ boson, or a $B$ or $D$ meson. 
The main systematic uncertainties arise from the modeling of the IP distribution for light jets in simulation, since a fit to this observable has been used to disentangle the different flavour components in the control samples prior to the SV-tagging requirement.
These corrections and uncertainties are verified in Ref.~\cite{top} to agree within 3\% between that data sample and the one used in this analysis. 
The systematic uncertainty for the SV-tagging efficiency, of around 20\%, dominates the total uncertainty on the cross-section measurements. It is correlated across all  bins of the analysis. Although the SV-tagging efficiencies are different for $b$- and $c$-jets, the SV-tagging systematic uncertainty is of the same order for the $b\bar{b}$- and $c\bar{c}$-dijet cross-sections, because it is related to the method described in Ref.~\cite{LHCb-PAPER-2015-016}, which affects in the same way the $b$- and $c$-jets efficiency determination. 
Uncertainties affecting the jet identification efficiency are evaluated as described in Ref.~\cite{LHCb-PAPER-2013-058}, using the \Zpjet data sample. The relative variation of the number of selected jets is compared between data and simulation, and the differences observed, which are at the level of 5\% per-jet, are used as a systematic uncertainty on $\epsilon_\text{reco}$. 

The uncertainty associated to differences between data and simulation in the jet energy resolution and jet energy scale affect the unfolding procedure, the acceptance factor and the efficiency measurement. Both the jet energy resolution and scale uncertainties are evaluated as explained in Ref.~\cite{LHCB-PAPER-2016-011}, using  the \Zpjet data sample introduced in Sec.~\ref{sec:selection}. To account for the jet energy resolution uncertainty, \Zpjet events  are used to evaluate the maximum gaussian smearing one needs to apply to the jet \pt in simulation to have an agreement with data within one standard deviation.
To determine the uncertainty arising from the jet energy scale, the same events are used to evaluate the multiplicative factor one needs to apply to the jet \pt in simulation to have an agreement within one standard deviation. The uncertainty associated with both of these effects is found to be negligible in this analysis.

Systematic uncertainties associated to the modelling of the templates may arise from  differences between data and simulation in the BDT classifiers distributions and affect the fitting procedure.
In order to evaluate them, the analysis is repeated using two other variables related to the SV employed for the jet SV-tagging: the corrected SV mass \cite{LHCb-PAPER-2015-016} and the number of tracks in the SV. In analogy with $t_0$ and $t_1$, new observables are built by summing these variables for $j_0$ and $j_1$, and the fits are then performed in the new space. The difference between these and the nominal results are used to evaluate the systematic uncertainties. These are on average at the level of 3\%. 
Concerning the fitting procedure itself, an alternative algorithm is applied, where rather than combining linearly, the responses of the two BDT classifiers are multiplied. Once again, the fits are repeated and the results compared with the nominal ones. The uncertainties for the $b\bar{b}$ yields are below 1\%, for the $c\bar{c}$ yields they are 13\% on average.

In order to assess the uncertainty due to the finite simulated samples size in the template construction, new templates are obtained with a ``bootstrapping with replacement'' technique \cite{bootstrap}. For each simulation sample used to build a template comprising $N$ events, an equal number of $N$ events are randomly extracted from the sample allowing to take multiple times the same event (repetitions).
This new set of events is then used to obtain a new template.
It has been demonstrated that the distribution of fit results obtained with the bootstrap technique mimics the distribution of results due to the finite simulated sample size~\cite{mcstat}.
The width of the distribution of the fit results using the different bootstrap templates is taken as the uncertainty associated to the simulated sample size.
The relative mean uncertainty is $0.8\%$ for $b \bar{b}$ and $3.6\%$ for $c \bar{c}$ yields. The finite simulated samples size also affects the efficiency evaluation, as well as the unfolding matrices. For the former, their effect is small compared to other uncertainties. For the latter, this is taken automatically into account by the unfolding algorithm \cite{tunfold}.  

The uncertainty on the modelling of initial-state (ISR) and final-state radiation (FSR) in simulation may affect the acceptance-factor determination, since different parametrisations of the gluon emission could change the jets kinematical distributions. Simulation samples where ISR and FSR parameters are varied are generated to determine the uncertainty. In particular the multiplicative factor applied to the renormalisation scale for ISR (FSR) branchings $\mu_{R}^{\text{ISR}}$ ($\mu_{R}^{\text{FSR}}$) is varied between 0.5 and 2, and the additive non-singular term in the ISR (FSR) splitting functions $c_{NS}^{\text{ISR}}$ ($c_{NS}^{\text{FSR}}$) is varied between $-2$ and 2 \cite{Sjostrand:2007gs,*Sjostrand:2006za}. The new acceptance factors are compared to the nominal ones, and their relative variation is considered as a systematic uncertainty. On average the variation is about 3\%. 

The unfolding matrix receives systematic uncertainties from all the different sources described in this section. These are propagated through the unfolding procedure. 
In the unfolding algorithm a regularisation parameter is chosen via a minimisation procedure. To further cross-check the algorithm, the regularisation parameter is varied around the minimum and the unfolding is repeated. Using a conservative approach the variation is chosen to be $\pm 50 \%$ of the parameter value. The difference with respect to the nominal result is taken as systematic uncertainty associated to the unfolding procedure, which is below 1\%. Another source of uncertainty is associated to the unfolding model. It is assessed by varying the underlying dijet kinematic distributions in the simulation samples used for the determination of the unfolding matrix. The unfolding procedure is repeated with this alternative set of unfolding matrices and unfolded distributions are compared with the nominal ones.
The relative variation in each bin is used as systematic uncertainty. This is again in average below 1\%.

Finally, the systematic uncertainty on the integrated luminosity is about $4\%$, determined as explained in Refs.~\cite{lumi, top}.

The systematic contributions from each source are summarized in Table \ref{tab:syst_results}. In the table, the mean relative uncertainties calculated averaging over $\eta(j_0)$ intervals are reported separately for $b\bar{b}$ and $c\bar{c}$ events. Since the different sources are considered to be uncorrelated, the total uncertainties are obtained by summing the individual uncertainties in quadrature. 
The unfolding-related systematic uncertainties are in principle correlated with some of the efficiency uncertainties, but since they are sub-dominant this correlation is neglected.
The dominant systematic uncertainties are those related to the GEC efficiency, the SV-tagging and the fit procedure.
\begin{table}[h]
  \begin{center}
      \caption{\label{tab:syst_results} Mean relative uncertainties in percent on the cross-sections calculated averaging over the leading jet pseudorapidity intervals for $b\bar{b}$ and $c\bar{c}$ events. For the total uncertainty individual contributions are added in quadrature.}
      \begin{tabular}{l|*2{S[table-format=<2.2, table-space-text-post=\,\%]}}
        \toprule
        Systematic source & {$\sigma(b\bar{b})$} & {$\sigma(c\bar{c})$}\\
        \midrule
        GEC &  6.3\,  &  6.3\,  \\
        Trigger & 4.2\, & 3.8\, \\
        Jet SV-tagging & 16\, & 20\, \\
        Jet identification & 4.5\, & 4.7\, \\
        Jet resolution & < 0.1\, & <0.1 \, \\
        Jet energy scale & < 0.1 \, & <0.1 \, \\
        Templates modelling & 2.8\, & 3.0\, \\
        Fit procedure &  0.4\, & 13\, \\
        Simulation sample size & 0.56\,  & 2.7\, \\
        Unfolding procedure & 0.20\, & 0.90\, \\
        Unfolding model & 0.17\, & 0.84\, \\
        ISR and FSR & 2.4\, & 2.4\, \\
        Luminosity & 3.8\, & 3.8\, \\
        \midrule
        Total & 19\, & 26\, \\ 
        \bottomrule
      \end{tabular}
  \end{center}
\end{table}
Finally, a closure test is performed to assess the validity of the analysis procedure. For this, the simulation samples are used to prepare a test dataset, and the full analysis chain is applied, from the fit to the unfolding. The measured and reference cross-sections are compatible within their statistical and systematic uncertainties.

\section{Ratio of $\mathbold{c \bar{c}}$- and $\mathbold{b \bar{b}}$-dijet cross-sections}
\label{sec:ratios}

This section presents the method used to determine the cross-section ratio, $R$, between \ccbar and \bbbar production. 
The measurement of $R$ is also performed in the different bins of kinematic observables: leading jet $\eta$, leading jet $p_{\mathrm{T}}$, $\Delta y^{*}$ and $m_{jj}$. The same binning scheme for reconstructed and generator-level observables is used as for the cross-section measurements.

Since several experimental and theoretical uncertainties cancel in the ratio it provides an excellent test of the SM and of pQCD, and can also be used to obtain valuable information when used in the global fits to extract the proton PDFs \cite{hadroproduction}.

 In analogy with Eq.~\ref{eq:cross}, the unfolded $R$ can be obtained with the following formula:
\begin{equation}
\label{eq:rcross}
R(i) = \frac{\sum\limits_{j=1}^{n} U_{ij} \frac{N^{c\bar{c}}(j)}{\epsilon^{c \bar{c}}_{\text{tag}}(j)}}{\sum\limits_{j=1}^{n} U_{ij} \frac{N^{b\bar{b}}(j)}{\epsilon^{b\bar{b}}_{\text{tag}}(j)}},
\end{equation}
where $i$ indicates the index of the bin defined for generator-level variables, $N^{c\bar{c}}(j)$ ($N^{b\bar{b}}(j)$) is the number of fitted $c \bar{c}$ ($b \bar{b}$) events in the bin $j$ defined for reconstructed variables,  $U_{ij}$ is the unfolding matrix introduced in Sec.~\ref{sec:unfolding} and $\epsilon_{\text{tag}}^{c \bar{c}}(j)$ ($\epsilon_{\text{tag}}
^{b \bar{b}}(j)$) is the $c \bar{c}$ ($b \bar{b}$) tagging efficiency in bin $j$. Apart from the tagging efficiency, which depends on the properties of $b$- and $c$-hadrons, it has been verified in simulation that all other efficiencies and acceptance factors are compatible and fully correlated between $b$- and $c$-jets, therefore they cancel in the ratio and they are not considered in the formula. 
The correlation between $N^{c\bar{c}}(j)$ and $N^{b\bar{b}(j)}$ is neglected when determining the uncertainty on $R$. This correlation leads to a small change on the statistical uncertainty and is negligible compared to the total uncertainty on $R$, which is in the order $20\%$.

Most sources of systematic uncertainty are common between the numerator and denominator of Eq.~\ref{eq:rcross}, cancelling their impact. The exceptions are the SV-tagging systematic  uncertainty, since this is measured on complementary data samples for $b$- and $c$-jets \cite{LHCb-PAPER-2015-016}; the fit procedure systematic uncertainty; the template modelling systematic uncertainty and the simulation sample size uncertainty.
These systematic uncertainties are considered uncorrelated. Although those related to the fit procedure should take into account the anti-correlations between $b\bar{b}$ and $c\bar{c}$ fitted yields, the systematic uncertainty introduced by neglecting such anti-correlations is found to be negligible with respect to the total uncertainty.


\section{Results and predictions}
\label{sec:results}

In this section the measurements of the $b\bar{b}$- and $c\bar{c}$-dijet differential cross-sections are presented, as well as the measurement of their ratio. 

The measurements in this section are compared with the pQCD NLO cross-section predictions obtained with \textsc{Madgraph5} aMC@NLO \cite{madgraph} for matrix elements computation and \pythia for parton showers. The predictions take into account the FSR and ISR contributions ~\cite{Sjostrand:2007gs,*Sjostrand:2006za}. 
The NNPDF2.3\_NLO set\cite{nnpdf23} has been used as PDF set for the calculation. At least two generator-level jets are required in the fiducial region, and the two jets with highest $p_{\mathrm{T}}$ that fulfill the requirements are used to calculate the differential distributions. The renormalisation ($\mu_r$) and factorisation scales ($\mu_f$) are set dynamically to the sum of transverse masses of all final-state particles divided by two.
The scale uncertainty has been obtained with an envelope of seven combinations of ($\mu_r$,$\mu_f$) values ($\mu_r,\mu_f=0.5,1,2$). The PDF uncertainty has been obtained as the envelope of 100 NNPDF2.3\_NLO replicas. 
The uncertainties on the predictions are correlated across the kinematical intervals considered for the measurement.
At high leading jet \pt and $m_{jj}$ the prediction uncertainties are of the order of 15$\%$, for both the $b\bar{b}$ and $c\bar{c}$ cross-sections. In principle more advanced techniques can reduce the prediction uncertainty \cite{hadroproduction} in the high $m_{jj}$ region, while phenomenological studies at low mass, where the renormalisation and factorisation scale uncertainty is larger, do not exist. 
The measurements are also compared with a leading-order prediction obtained with \pythia for both process generation and parton showering.

\begin{figure}[htb]
  \centering
    \includegraphics[width=0.49\linewidth]{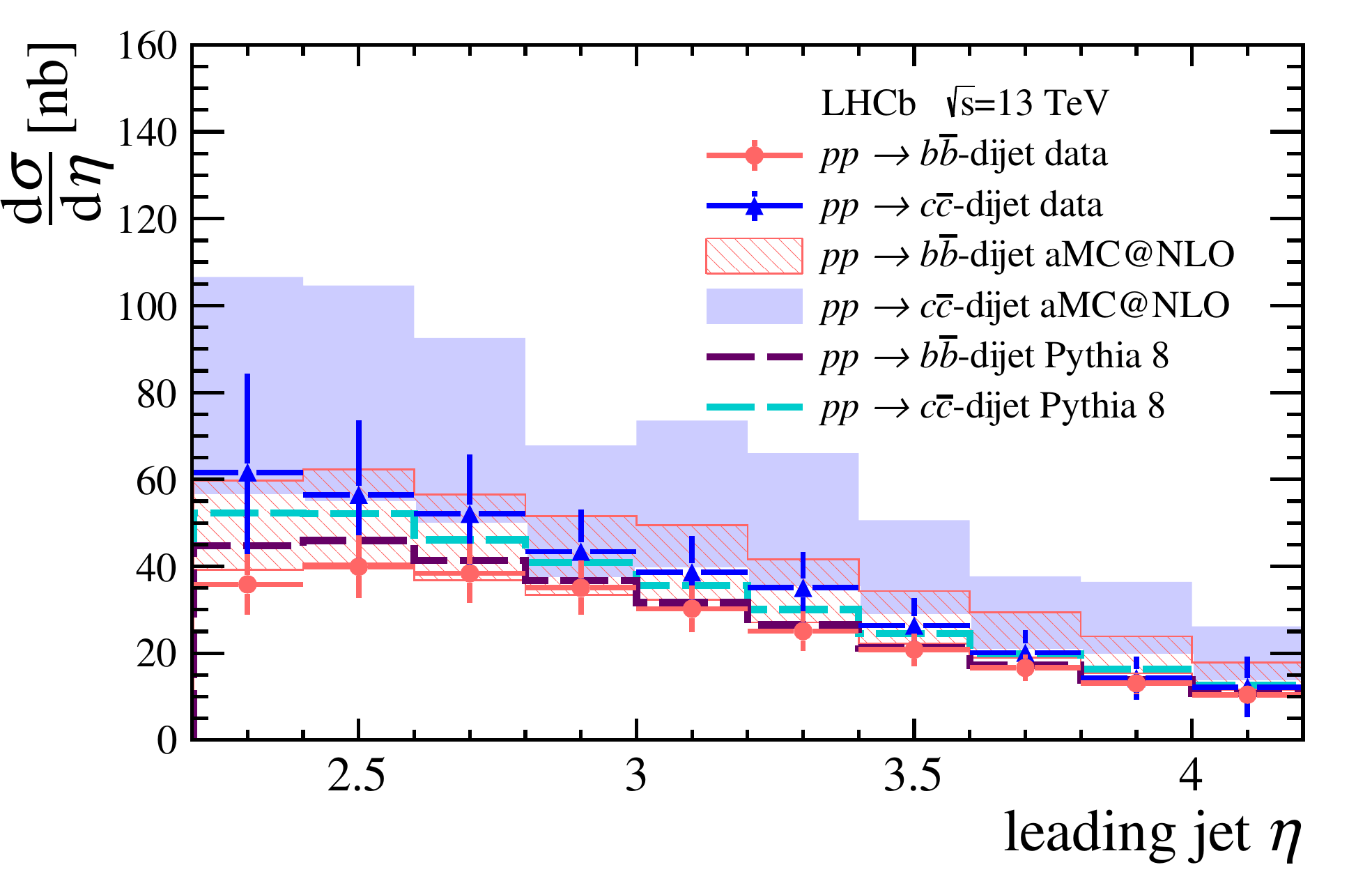}
    \includegraphics[width=0.49\linewidth]{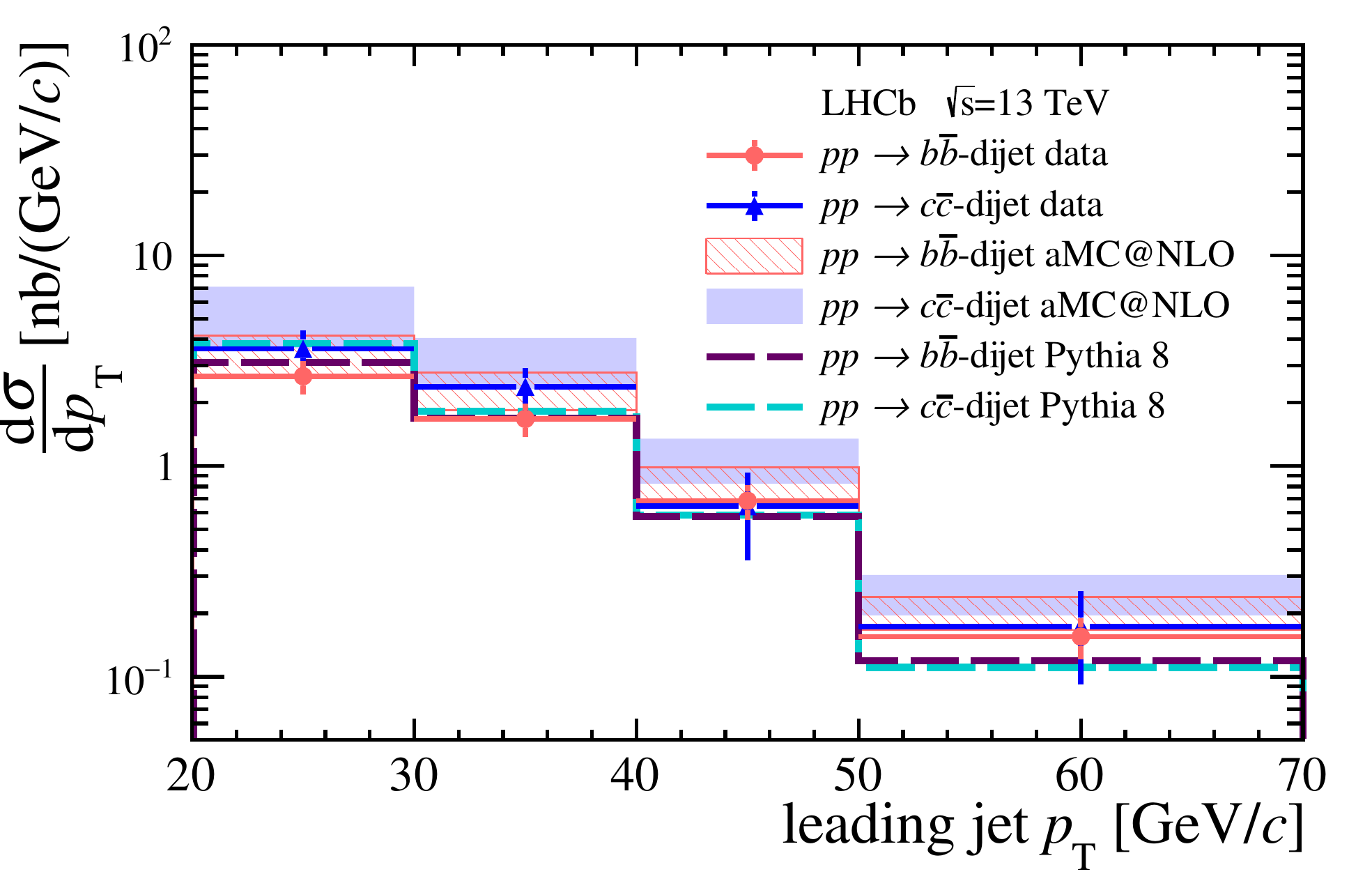} \\
    \includegraphics[width=0.49\linewidth]{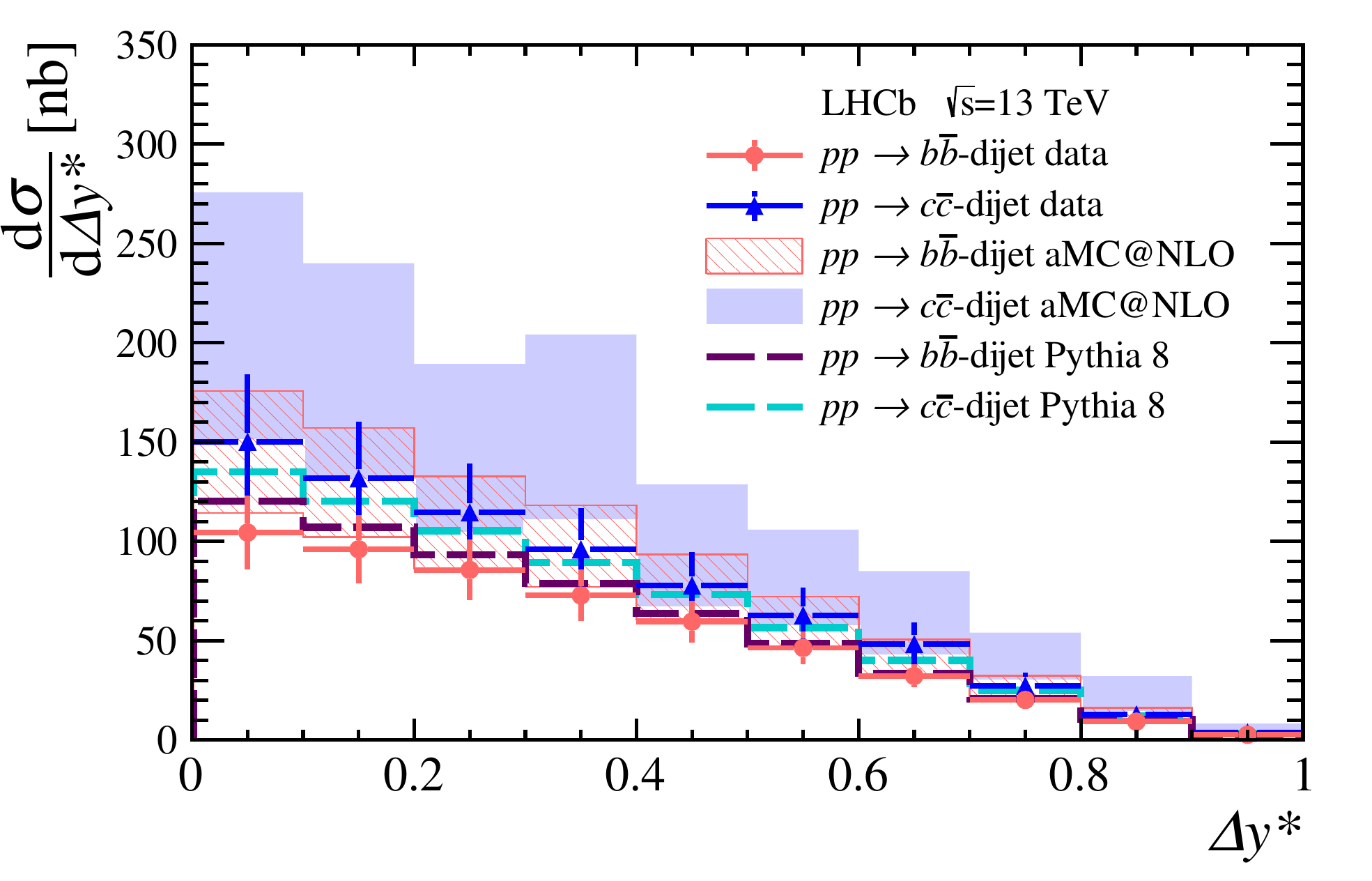}
    \includegraphics[width=0.49\linewidth]{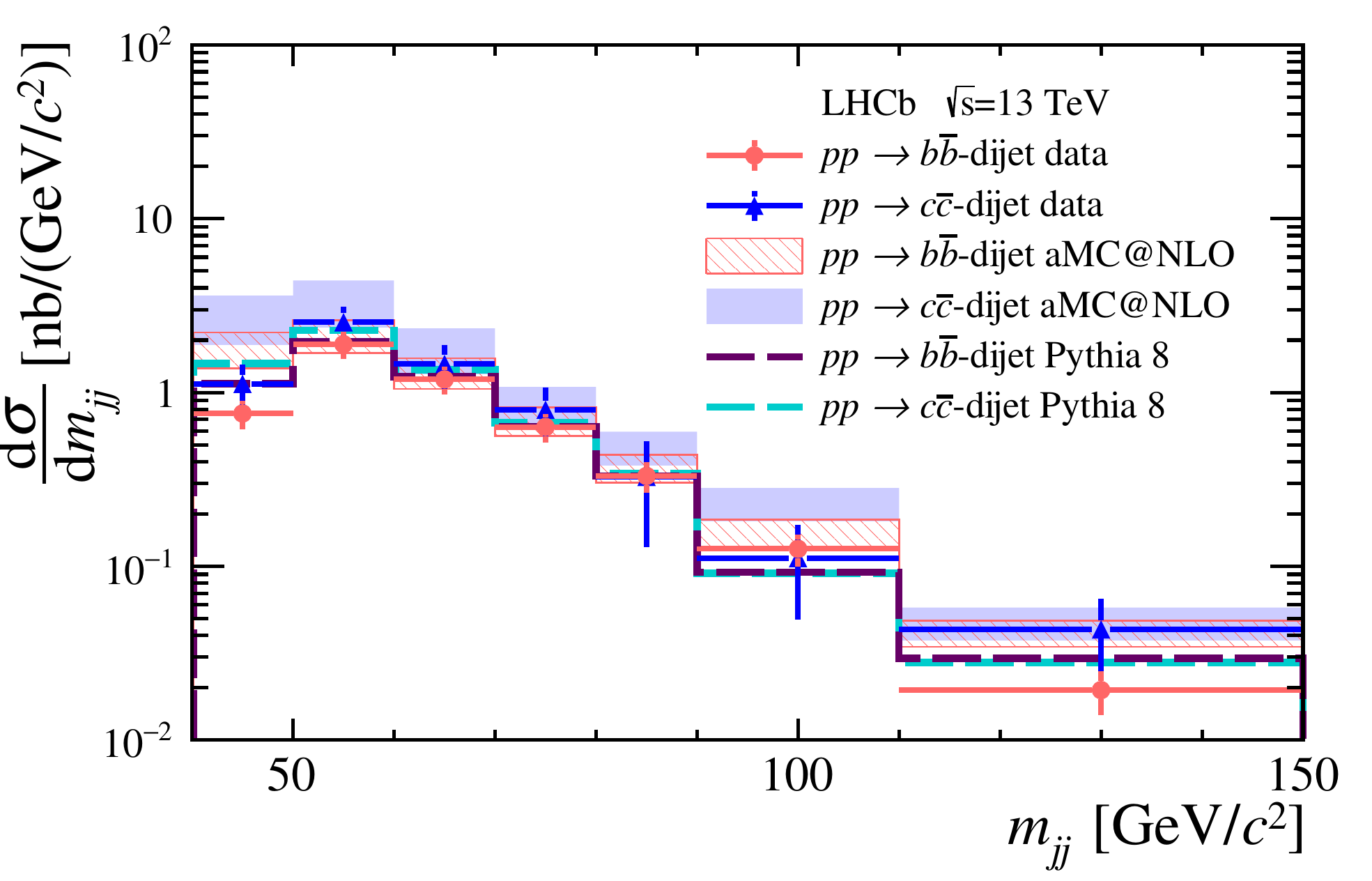}
  \caption{Differential $b\bar{b}$- and $c\bar{c}$-dijet cross-sections as a function of the (top left) leading jet $\eta$, (top right) the leading jet $p_{\mathrm{T}}$, (bottom left) $\Delta y^*$ and (bottom right) $m_{jj}$. The error bars represent the total uncertainties, that are almost fully correlated across the bins. The next-to-leading-order predictions obtained with Madgraph5 aMC@NLO + \pythia are shown. The prediction uncertainty is dominated by the renormalisation and factorisation scale uncertainty. The leading-order prediction obtained with \pythia is also shown.
    \label{fig:xs}}
\end{figure}

Figure~\ref{fig:xs} shows the $b\bar{b}$- and $c \bar{c}$-dijet differential cross-sections as a function of the leading jet $\eta$, the leading jet $p_{\mathrm{T}}$, $\Delta y^{*}$ and $m_{jj}$.
The cross-sections as a function of $\Delta y^{*}$ and $m_{jj}$ are presented in logarithmic scale, while in App.~\ref{app:linear} they are presented in linear scale.
The numerical values of the measured cross-sections, the covariance matrices for the $b\bar{b}$ ($c\bar{c}$) intervals and the cross-correlation matrix between $b\bar{b}$ and $c \bar{c}$ intervals are reported in App.~\ref{app:cov}. 
The total uncertainty is almost fully correlated across the bins, since it is dominated by common systematic uncertainties. The only uncorrelated contributions to the total uncertainty are the statistical uncertainty and the systematic uncertainty related to the finite simulated sample size, which are negligible with respect to the total uncertainty.
Note that the leading jet $p_{\mathrm{T}}$ and $m_{jj}$ ranges are reduced to [20,70]\gevc and [40,150]\gevc respectively, because the unfolding produces cross-sections compatible with zero events in the high $p_{\mathrm{T}}$ and high mass bins. The measurements are generally slightly below the predictions.
The compatibility of the measurements with the prediction, obtained including the uncertainties on both, is within 1 to 2 standard deviations. It can be noticed that the predictions at low leading jet \pt and $m_{jj}$ show large uncertainties, that are dominated by the renormalisation and factorisation scale uncertainty.
The global compatibility of the measurements with predictions, calculated considering the correlations between the different bins, is 0.9\,$\sigma$ for the $b\bar{b}$- and 0.8\,$\sigma$ for the $c\bar{c}$-dijet cross-sections.

\begin{figure}[htb]
  \centering
    \includegraphics[width=0.49\linewidth]{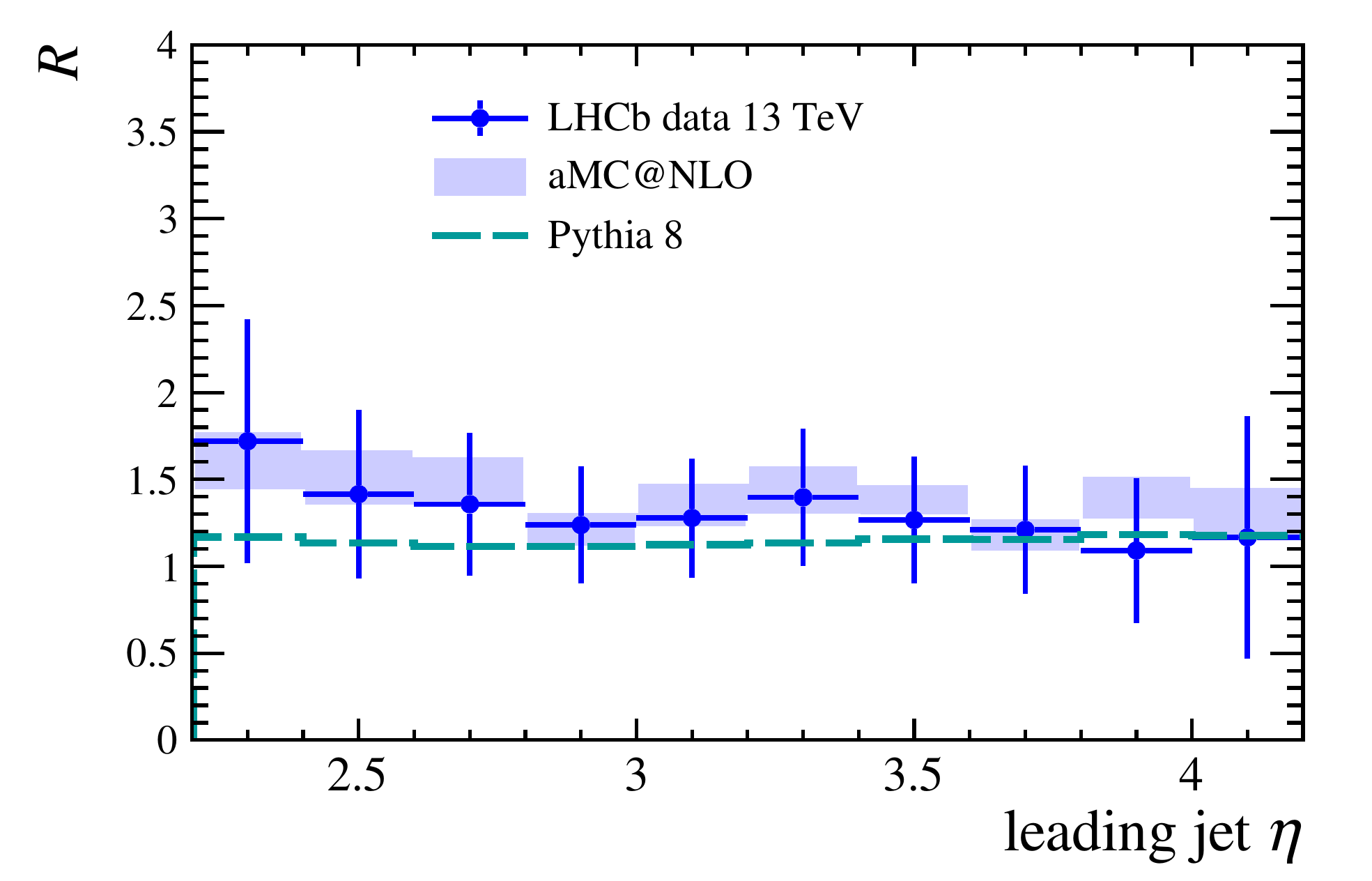}
    \includegraphics[width=0.49\linewidth]{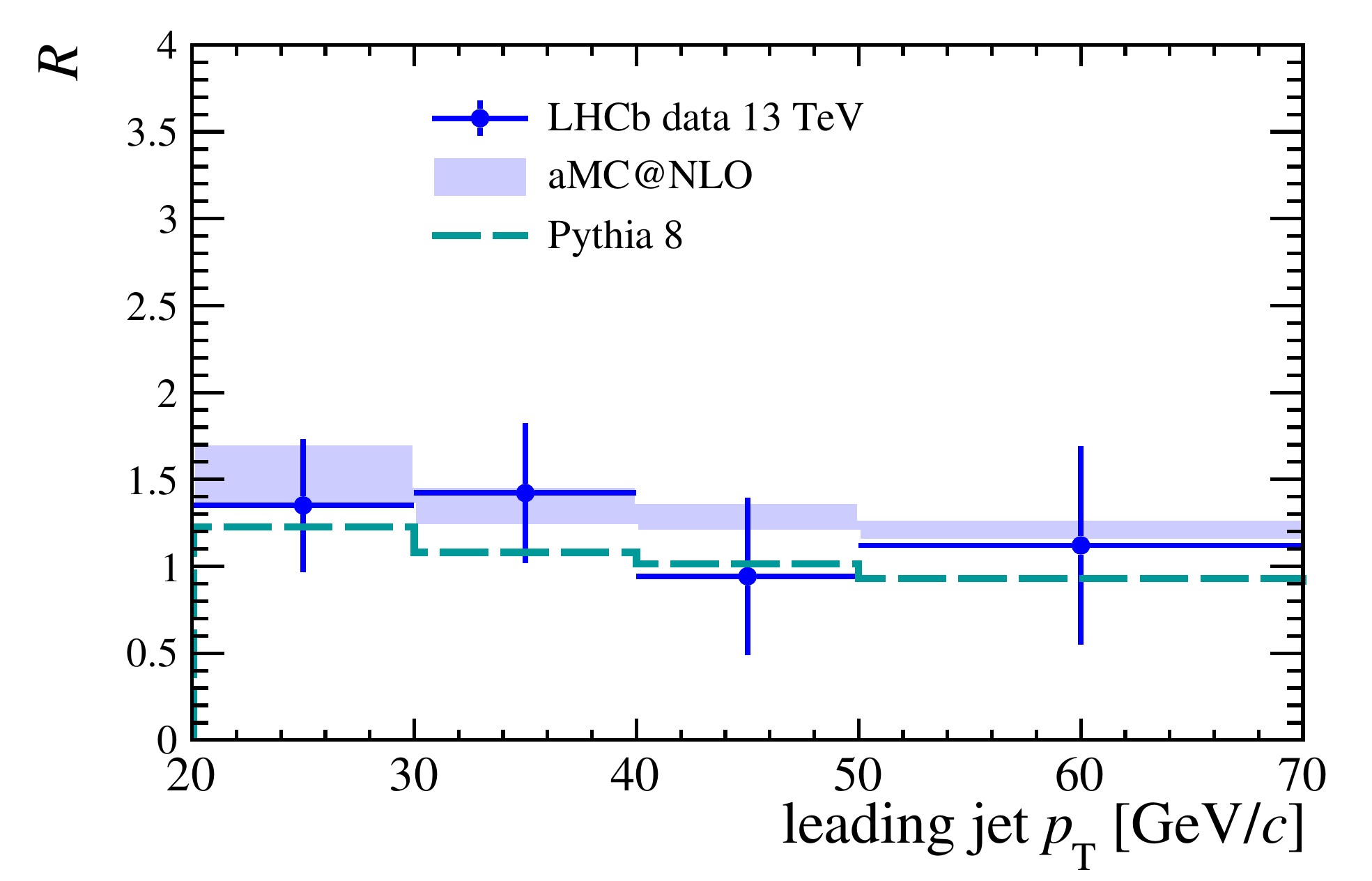} \\
    \includegraphics[width=0.49\linewidth]{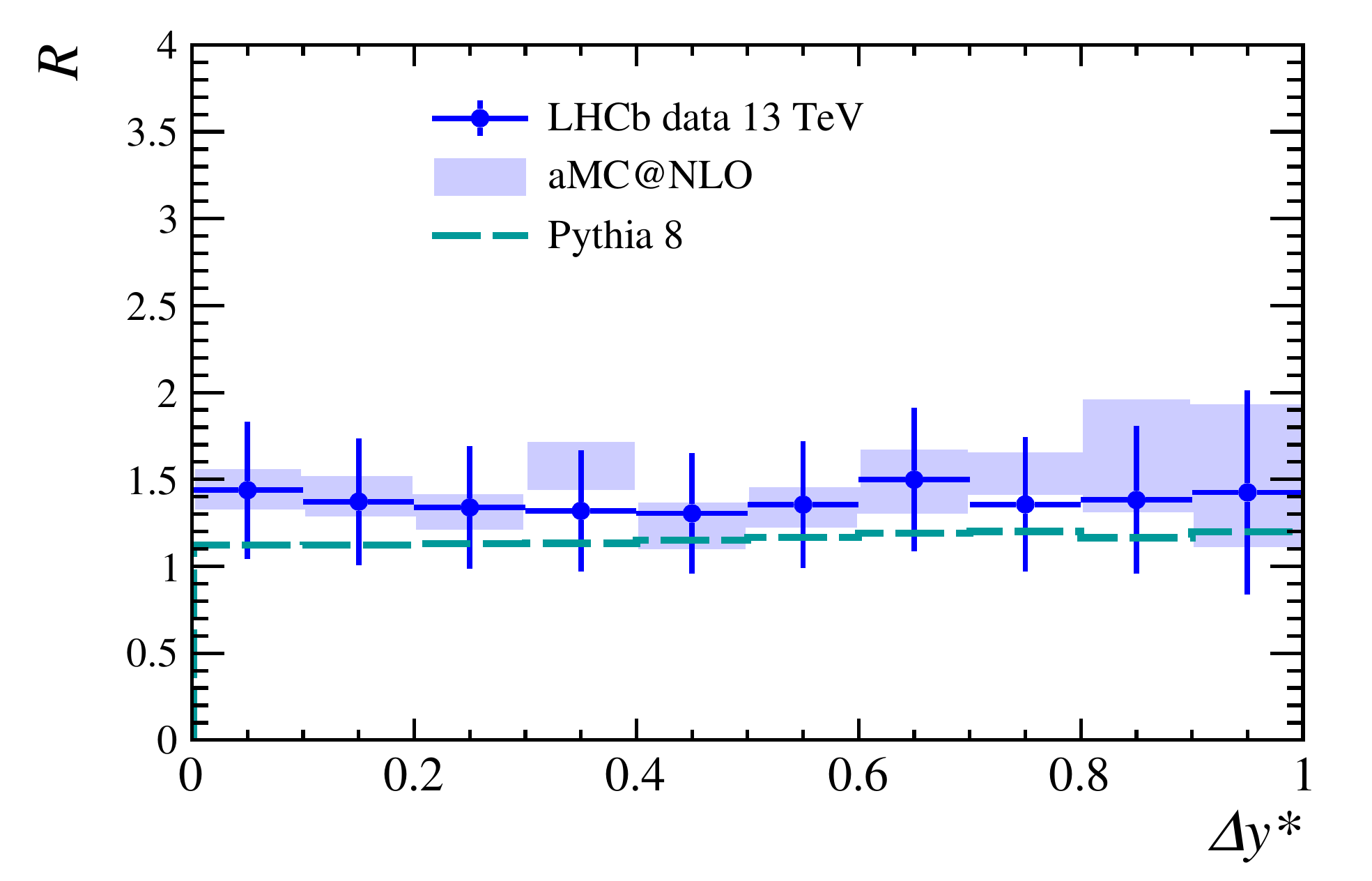}
    \includegraphics[width=0.49\linewidth]{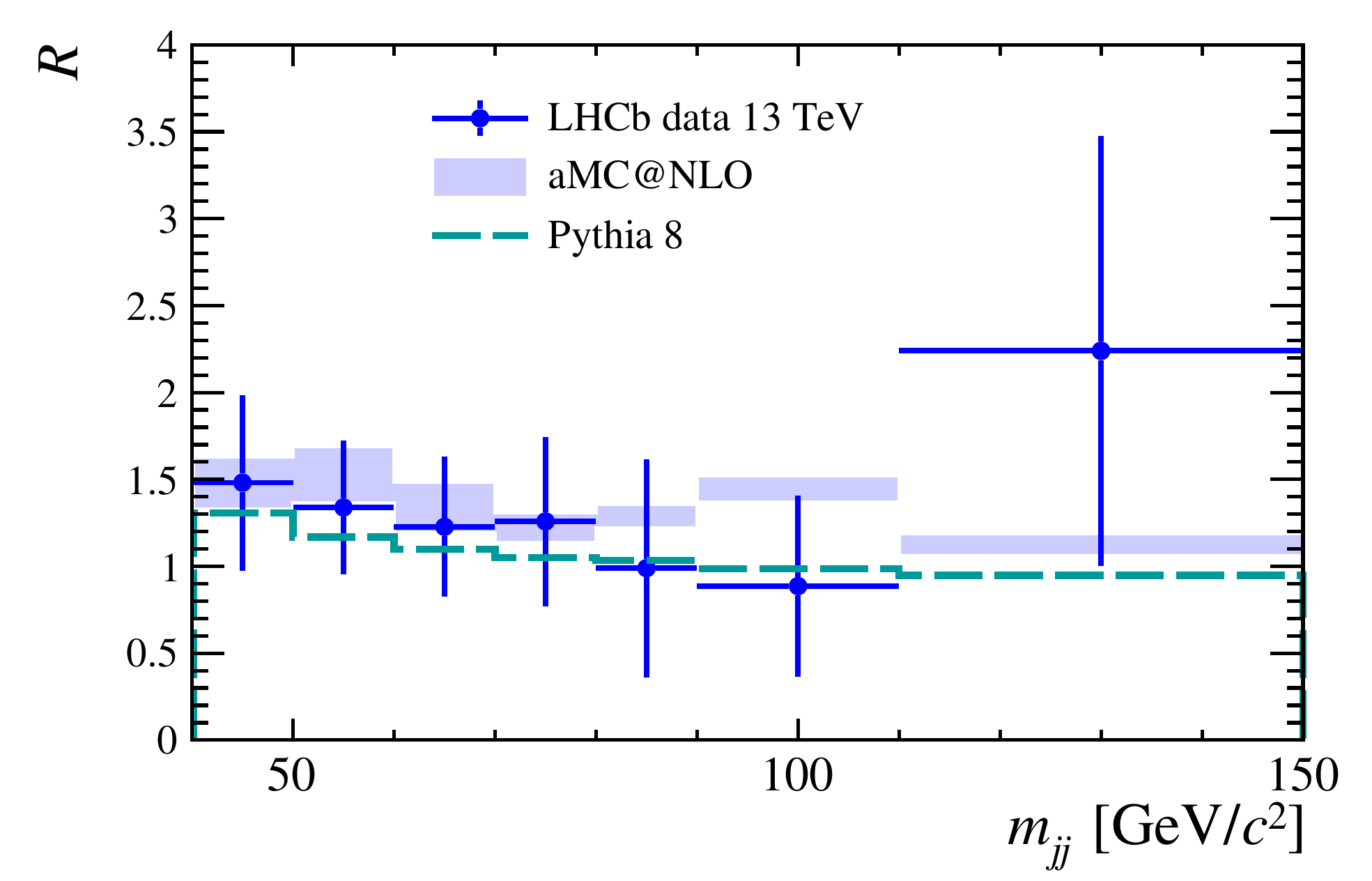}
  \caption{ Measured $c\bar{c}$ to $b \bar{b}$ cross-section ratio as a function of (top left) the leading jet $\eta$, (top right) the leading jet $p_{\mathrm{T}}$, (bottom left) $\Delta y^*$  and (bottom right) $m_{jj}$. The error bars represent the total uncertainties, that are almost fully correlated across the bins. The next-to-leading-order predictions obtained with Madgraph5 aMC@NLO + \pythia are shown. The prediction uncertainty is dominated by the renormalisation and factorisation scale uncertainty. The leading-order prediction obtained with \pythia is also shown.
    \label{fig:R}}
\end{figure}

Figure \ref{fig:R} shows the cross-section ratios $R$ as a function of the leading jet $\eta$, the leading jet $p_{\mathrm{T}}$, $\Delta y^{*}$ and $m_{jj}$. The $R$ measurements are compatible with the prediction within its uncertainties. It can be noticed that the measured ratio $R$ is in the order of 1.4, significantly lower than the inclusive $c\bar{c}$/$b\bar{b}$ ratio expected in $pp$ collisions: this is due to the jet $p_{\mathrm{T}}>$ 20 \gev requirement of the fiducial region that partially compensates the effect of the different $b$- and $c$-quark masses.


The differential distributions are summed up to obtain the integrated cross-sections in the fiducial region.
In this way four different values of $\sigma(pp \rightarrow b \bar{b}\text{-dijet} \, X)$, $\sigma(pp \rightarrow c \bar{c}\text{-dijet} \, X)$, where $X$ indicates additional particles produced in the collisions, and $R$ are obtained, one for each observable. 
The different measurements of the same quantity are in agreement within their total uncertainty. The integrated measurements obtained from the $\Delta y^*$ distributions have the smallest relative uncertainty from the fit procedure and are considered as the nominal result.
The total integrated $b \bar{b}$- and $c \bar{c}$-dijet cross-sections and their ratio are presented in Table \ref{tab:xs_tot}. In this table the statistical, systematic and luminosity uncertainties are presented separately. The total cross-sections and $R$ are compatible with the prediction from \textsc{Madgraph5} aMC@NLO + \pythia within its uncertainty.
\begin{table}[b]
  \begin{center}
      \caption{\label{tab:xs_tot} The total $b \bar{b}$-dijet and $c \bar{c}$-dijet cross-sections and their ratio in the fiducial region, compared with the NLO predictions. The first uncertainty on the measurement is the combined statistical and systematic uncertainty and the second is the uncertainty from the luminosity. For the measurement of $R$ the luminosity uncertainty cancels in the ratio. The statistical uncertainty for the cross-section and $R$ measurements is also reported. For the predictions the first uncertainty corresponds to the scale uncertainty, the second to the PDF uncertainty.}
      \bgroup
       \def\arraystretch{1.75}%
      \begin{tabular}{c|p{.5cm}p{.8cm}p{.9cm}|c|p{.35cm}p{.4cm}p{.9cm}}
      \toprule
        observable & \multicolumn{3}{c|}{measurement} & stat.~uncertainty & \multicolumn{3}{c}{prediction}   \\
        \midrule
        $\sigma(b \bar{b})$ [nb]  & 53.0 & $\pm~\phantom{0}9.5$ & $ \pm~2.1 $ & $< 0.1$ & $70.2$ & \phantom{}$^{+15.1}_{-14.7}$ & $ \phantom{}^{+1.4}_{-1.4}$ \\
        $\sigma(c \bar{c})$ [nb] & 72.6 & $\pm~16.1$ & $\pm~2.9$ & $< 0.1$ & $97.9$ & \phantom{}$^{+34.5}_{-27.5}$ & $\phantom{}^{+1.8}_{-1.8}$ \\
        $R$ & 1.37 & $\pm~0.27$ & & $< 0.01 $ & $1.39$ & \phantom{}$^{+0.16}_{-0.13}$ & $\phantom{}^{+0.03}_{-0.03}$ \\
        \bottomrule 
      \end{tabular}
      \egroup
  \end{center}
\end{table}


\section{Summary}
\label{sec:Conclusions}
 
Measurements of the total and differential $b \bar{b}$- and $c \bar{c}$-dijet production cross-sections in $pp$ collisions at $\sqrt{s}=13$ TeV in the \lhcb acceptance have been presented. The ratio, $R$, between the $c \bar{c}$- and $b \bar{b}$-dijet cross-sections has also been measured. Results are presented for the fiducial region for generator-level jets with transverse momentum $p_{\mathrm{T}}>20$ \gevc, pseudorapidity $2.2 <  \eta < 4.2$ and azimuthal difference $|\Delta \phi|>1.5$.

The total measured $b \bar{b}$-dijet cross-section in the fiducial region is
\begin{displaymath}
\sigma(pp \rightarrow b \bar{b}\text{-dijet} \, X) = 53.0 \pm 9.5 \pm 2.1 ~ \mathrm{nb},
\end{displaymath}
where the first uncertainty is the combined statistical and systematic uncertainty and the second is due to the precision of the luminosity calibration.
The statistical uncertainty is small, corresponding to $0.012\%$ of the measured $b\bar{b}$-dijet cross-section value.
The total measured $c \bar{c}$-dijet cross-section in the fiducial region is
\begin{displaymath}
\sigma(pp \rightarrow c \bar{c}\text{-dijet} \, X) = 72.6 \pm 16.1 \pm 2.9 ~ \mathrm{nb},
\end{displaymath}
and the statistical uncertainty corresponds to $0.03\%$ of the measured $c\bar{c}$-dijet cross-section value.
The measured ratio between the two cross-sections is 
\begin{displaymath}
R=\frac{\sigma(pp \rightarrow c \bar{c}\text{-dijet} \, X)}{\sigma(pp \rightarrow b \bar{b}\text{-dijet} X)} \, = 1.37 \pm 0.27,
\end{displaymath}
and the statistical uncertainty corresponds to $0.03\%$ of the measured cross-sections ratio value.
The total cross-sections and the ratio between the two are compatible with the Madgraph5 aMC@NLO + \pythia expectation within the total uncertainties.  Differential cross-sections are measured as a function of  the leading jet $\eta$,  the leading jet $p_{\mathrm{T}}$,  $\Delta y^*$  and $m_{jj}$ and found to agree within 1 to 2 standard deviations with the predictions, depending on the intervals. The numerical values of the cross-sections and cross-sections ratios are summarized in App.~\ref{app:cov}.
This is the first inclusive, direct measurement of the differential $c \bar{c}$-dijet production cross-section at a hadron collider.

\clearpage
\section*{Acknowledgements}
%
%
\noindent We express our gratitude to our colleagues in the CERN
accelerator departments for the excellent performance of the LHC. We
thank the technical and administrative staff at the LHCb
institutes.
We acknowledge support from CERN and from the national agencies:
CAPES, CNPq, FAPERJ and FINEP (Brazil); 
MOST and NSFC (China); 
CNRS/IN2P3 (France); 
BMBF, DFG and MPG (Germany); 
INFN (Italy); 
NWO (Netherlands); 
MNiSW and NCN (Poland); 
MEN/IFA (Romania); 
MSHE (Russia); 
MICINN (Spain); 
SNSF and SER (Switzerland); 
NASU (Ukraine); 
STFC (United Kingdom); 
DOE NP and NSF (USA).
We acknowledge the computing resources that are provided by CERN, IN2P3
(France), KIT and DESY (Germany), INFN (Italy), SURF (Netherlands),
PIC (Spain), GridPP (United Kingdom), RRCKI and Yandex
LLC (Russia), CSCS (Switzerland), IFIN-HH (Romania), CBPF (Brazil),
PL-GRID (Poland) and OSC (USA).
We are indebted to the communities behind the multiple open-source
software packages on which we depend.
Individual groups or members have received support from
AvH Foundation (Germany);
EPLANET, Marie Sk\l{}odowska-Curie Actions and ERC (European Union);
A*MIDEX, ANR, Labex P2IO and OCEVU, and R\'{e}gion Auvergne-Rh\^{o}ne-Alpes (France);
Key Research Program of Frontier Sciences of CAS, CAS PIFI,
Thousand Talents Program, and Sci. \& Tech. Program of Guangzhou (China);
RFBR, RSF and Yandex LLC (Russia);
GVA, XuntaGal and GENCAT (Spain);
the Royal Society
and the Leverhulme Trust (United Kingdom).

\clearpage


\addcontentsline{toc}{section}{References}
\bibliographystyle{LHCb}
\bibliography{main,standard,LHCb-PAPER,LHCb-CONF,LHCb-DP,LHCb-TDR}

\ifx\mcitethebibliography\mciteundefinedmacro
\PackageError{LHCb.bst}{mciteplus.sty has not been loaded}
{This bibstyle requires the use of the mciteplus package.}\fi
\providecommand{\href}[2]{#2}
\begin{mcitethebibliography}{10}
\mciteSetBstSublistMode{n}
\mciteSetBstMaxWidthForm{subitem}{\alph{mcitesubitemcount})}
\mciteSetBstSublistLabelBeginEnd{\mcitemaxwidthsubitemform\space}
{\relax}{\relax}

\bibitem{Cacciari:2012ny}
M.~Cacciari {\em et~al.},
  \ifthenelse{\boolean{articletitles}}{\emph{{Theoretical predictions for charm
  and bottom production at the LHC}},
  }{}\href{https://doi.org/10.1007/JHEP10(2012)137}{JHEP \textbf{10} (2012)
  137}, \href{http://arxiv.org/abs/1205.6344}{{\normalfont\ttfamily
  arXiv:1205.6344}}\relax
\mciteBstWouldAddEndPuncttrue
\mciteSetBstMidEndSepPunct{\mcitedefaultmidpunct}
{\mcitedefaultendpunct}{\mcitedefaultseppunct}\relax
\EndOfBibitem
\bibitem{Kniehl:2011bk}
B.~A. Kniehl, G.~Kramer, I.~Schienbein, and H.~Spiesberger,
  \ifthenelse{\boolean{articletitles}}{\emph{{Inclusive $B$-meson production at
  the LHC in the GM-VFN scheme}},
  }{}\href{https://doi.org/10.1103/PhysRevD.84.094026}{Phys.\ Rev.\
  \textbf{D84} (2011) 094026},
  \href{http://arxiv.org/abs/1109.2472}{{\normalfont\ttfamily
  arXiv:1109.2472}}\relax
\mciteBstWouldAddEndPuncttrue
\mciteSetBstMidEndSepPunct{\mcitedefaultmidpunct}
{\mcitedefaultendpunct}{\mcitedefaultseppunct}\relax
\EndOfBibitem
\bibitem{madgraph}
J.~Alwall {\em et~al.}, \ifthenelse{\boolean{articletitles}}{\emph{{The
  automated computation of tree-level and next-to-leading order differential
  cross sections, and their matching to parton shower simulations}},
  }{}\href{https://doi.org/10.1007/JHEP07(2014)079}{JHEP \textbf{07} (2014)
  079}, \href{http://arxiv.org/abs/1405.0301}{{\normalfont\ttfamily
  arXiv:1405.0301}}\relax
\mciteBstWouldAddEndPuncttrue
\mciteSetBstMidEndSepPunct{\mcitedefaultmidpunct}
{\mcitedefaultendpunct}{\mcitedefaultseppunct}\relax
\EndOfBibitem
\bibitem{hadroproduction}
R.~Gauld, U.~Haisch, and B.~D. Pecjak,
  \ifthenelse{\boolean{articletitles}}{\emph{{Asymmetric heavy-quark
  hadroproduction at LHCb: Predictions and applications.}},
  }{}\href{https://doi.org/10.1007/JHEP03(2019)166}{JHEP \textbf{03} (2019)
  166}, \href{http://arxiv.org/abs/1901.07573}{{\normalfont\ttfamily
  arXiv:1901.07573}}\relax
\mciteBstWouldAddEndPuncttrue
\mciteSetBstMidEndSepPunct{\mcitedefaultmidpunct}
{\mcitedefaultendpunct}{\mcitedefaultseppunct}\relax
\EndOfBibitem
\bibitem{Rojo:2017xpe}
J.~Rojo, \ifthenelse{\boolean{articletitles}}{\emph{{Improving quark flavor
  separation with forward W and Z production at LHCb}},
  }{}\href{https://doi.org/10.22323/1.297.0198}{PoS \textbf{DIS2017} (2018)
  198}, \href{http://arxiv.org/abs/1705.04468}{{\normalfont\ttfamily
  arXiv:1705.04468}}\relax
\mciteBstWouldAddEndPuncttrue
\mciteSetBstMidEndSepPunct{\mcitedefaultmidpunct}
{\mcitedefaultendpunct}{\mcitedefaultseppunct}\relax
\EndOfBibitem
\bibitem{LHCB-CONF-2013-002}
{LHCb collaboration}, \ifthenelse{\boolean{articletitles}}{\emph{{Measurement
  of $\sigma(b\bquarkbar)$ with inclusive final states}}, }{}
  \href{http://cdsweb.cern.ch/search?p=LHCb-CONF-2013-002&f=reportnumber&action_search=Search&c=LHCb+Conference+Contributions}
  {LHCb-CONF-2013-002}, {2013}\relax
\mciteBstWouldAddEndPuncttrue
\mciteSetBstMidEndSepPunct{\mcitedefaultmidpunct}
{\mcitedefaultendpunct}{\mcitedefaultseppunct}\relax
\EndOfBibitem
\bibitem{LHCb-PAPER-2016-031}
LHCb collaboration, R.~Aaij {\em et~al.},
  \ifthenelse{\boolean{articletitles}}{\emph{{Measurement of the
  $\bquark$-quark production cross-section in 7 and 13\tev $\proton\proton$
  collisions}}, }{}\href{https://doi.org/10.1103/PhysRevLett.118.052002}{Phys.\
  Rev.\ Lett.\  \textbf{118} (2017) 052002}, Erratum
  \href{https://doi.org/10.1103/PhysRevLett.119.169901}{ibid.\   \textbf{119}
  (2017) 169901}, \href{http://arxiv.org/abs/1612.05140}{{\normalfont\ttfamily
  arXiv:1612.05140}}\relax
\mciteBstWouldAddEndPuncttrue
\mciteSetBstMidEndSepPunct{\mcitedefaultmidpunct}
{\mcitedefaultendpunct}{\mcitedefaultseppunct}\relax
\EndOfBibitem
\bibitem{Aaboud:2016jed}
ATLAS collaboration, M.~Aaboud {\em et~al.},
  \ifthenelse{\boolean{articletitles}}{\emph{{Measurement of the
  $b\overline{b}$ dijet cross section in pp collisions at $\sqrt{s} = 7$ TeV
  with the ATLAS detector}},
  }{}\href{https://doi.org/10.1140/epjc/s10052-016-4521-y}{Eur.\ Phys.\ J.\
  \textbf{C76} (2016) 670},
  \href{http://arxiv.org/abs/1607.08430}{{\normalfont\ttfamily
  arXiv:1607.08430}}\relax
\mciteBstWouldAddEndPuncttrue
\mciteSetBstMidEndSepPunct{\mcitedefaultmidpunct}
{\mcitedefaultendpunct}{\mcitedefaultseppunct}\relax
\EndOfBibitem
\bibitem{Chatrchyan:2012dk}
CMS collaboration, S.~Chatrchyan {\em et~al.},
  \ifthenelse{\boolean{articletitles}}{\emph{{Inclusive $b$-jet production in
  $pp$ collisions at $\sqrt{s}=7$ TeV}},
  }{}\href{https://doi.org/10.1007/JHEP04(2012)084}{JHEP \textbf{04} (2012)
  084}, \href{http://arxiv.org/abs/1202.4617}{{\normalfont\ttfamily
  arXiv:1202.4617}}\relax
\mciteBstWouldAddEndPuncttrue
\mciteSetBstMidEndSepPunct{\mcitedefaultmidpunct}
{\mcitedefaultendpunct}{\mcitedefaultseppunct}\relax
\EndOfBibitem
\bibitem{Alves:2008zz}
LHCb collaboration, A.~A. Alves~Jr.\ {\em et~al.},
  \ifthenelse{\boolean{articletitles}}{\emph{{The \lhcb detector at the LHC}},
  }{}\href{https://doi.org/10.1088/1748-0221/3/08/S08005}{JINST \textbf{3}
  (2008) S08005}\relax
\mciteBstWouldAddEndPuncttrue
\mciteSetBstMidEndSepPunct{\mcitedefaultmidpunct}
{\mcitedefaultendpunct}{\mcitedefaultseppunct}\relax
\EndOfBibitem
\bibitem{LHCb-DP-2014-002}
LHCb collaboration, R.~Aaij {\em et~al.},
  \ifthenelse{\boolean{articletitles}}{\emph{{LHCb detector performance}},
  }{}\href{https://doi.org/10.1142/S0217751X15300227}{Int.\ J.\ Mod.\ Phys.\
  \textbf{A30} (2015) 1530022},
  \href{http://arxiv.org/abs/1412.6352}{{\normalfont\ttfamily
  arXiv:1412.6352}}\relax
\mciteBstWouldAddEndPuncttrue
\mciteSetBstMidEndSepPunct{\mcitedefaultmidpunct}
{\mcitedefaultendpunct}{\mcitedefaultseppunct}\relax
\EndOfBibitem
\bibitem{LHCb-DP-2014-001}
R.~Aaij {\em et~al.}, \ifthenelse{\boolean{articletitles}}{\emph{{Performance
  of the LHCb Vertex Locator}},
  }{}\href{https://doi.org/10.1088/1748-0221/9/09/P09007}{JINST \textbf{9}
  (2014) P09007}, \href{http://arxiv.org/abs/1405.7808}{{\normalfont\ttfamily
  arXiv:1405.7808}}\relax
\mciteBstWouldAddEndPuncttrue
\mciteSetBstMidEndSepPunct{\mcitedefaultmidpunct}
{\mcitedefaultendpunct}{\mcitedefaultseppunct}\relax
\EndOfBibitem
\bibitem{LHCb-DP-2017-001}
P.~d'Argent {\em et~al.}, \ifthenelse{\boolean{articletitles}}{\emph{{Improved
  performance of the LHCb Outer Tracker in LHC Run 2}},
  }{}\href{https://doi.org/10.1088/1748-0221/12/11/P11016}{JINST \textbf{9}
  (2017) P11016}, \href{http://arxiv.org/abs/1708.00819}{{\normalfont\ttfamily
  arXiv:1708.00819}}\relax
\mciteBstWouldAddEndPuncttrue
\mciteSetBstMidEndSepPunct{\mcitedefaultmidpunct}
{\mcitedefaultendpunct}{\mcitedefaultseppunct}\relax
\EndOfBibitem
\bibitem{LHCb-DP-2012-003}
M.~Adinolfi {\em et~al.},
  \ifthenelse{\boolean{articletitles}}{\emph{{Performance of the \lhcb RICH
  detector at the LHC}},
  }{}\href{https://doi.org/10.1140/epjc/s10052-013-2431-9}{Eur.\ Phys.\ J.\
  \textbf{C73} (2013) 2431},
  \href{http://arxiv.org/abs/1211.6759}{{\normalfont\ttfamily
  arXiv:1211.6759}}\relax
\mciteBstWouldAddEndPuncttrue
\mciteSetBstMidEndSepPunct{\mcitedefaultmidpunct}
{\mcitedefaultendpunct}{\mcitedefaultseppunct}\relax
\EndOfBibitem
\bibitem{LHCb-DP-2012-002}
A.~A. Alves~Jr.\ {\em et~al.},
  \ifthenelse{\boolean{articletitles}}{\emph{{Performance of the LHCb muon
  system}}, }{}\href{https://doi.org/10.1088/1748-0221/8/02/P02022}{JINST
  \textbf{8} (2013) P02022},
  \href{http://arxiv.org/abs/1211.1346}{{\normalfont\ttfamily
  arXiv:1211.1346}}\relax
\mciteBstWouldAddEndPuncttrue
\mciteSetBstMidEndSepPunct{\mcitedefaultmidpunct}
{\mcitedefaultendpunct}{\mcitedefaultseppunct}\relax
\EndOfBibitem
\bibitem{LHCb-DP-2012-004}
R.~Aaij {\em et~al.}, \ifthenelse{\boolean{articletitles}}{\emph{{The \lhcb
  trigger and its performance in 2011}},
  }{}\href{https://doi.org/10.1088/1748-0221/8/04/P04022}{JINST \textbf{8}
  (2013) P04022}, \href{http://arxiv.org/abs/1211.3055}{{\normalfont\ttfamily
  arXiv:1211.3055}}\relax
\mciteBstWouldAddEndPuncttrue
\mciteSetBstMidEndSepPunct{\mcitedefaultmidpunct}
{\mcitedefaultendpunct}{\mcitedefaultseppunct}\relax
\EndOfBibitem
\bibitem{Sjostrand:2007gs}
T.~Sj\"{o}strand, S.~Mrenna, and P.~Skands,
  \ifthenelse{\boolean{articletitles}}{\emph{{A brief introduction to PYTHIA
  8.1}}, }{}\href{https://doi.org/10.1016/j.cpc.2008.01.036}{Comput.\ Phys.\
  Commun.\  \textbf{178} (2008) 852},
  \href{http://arxiv.org/abs/0710.3820}{{\normalfont\ttfamily
  arXiv:0710.3820}}\relax
\mciteBstWouldAddEndPuncttrue
\mciteSetBstMidEndSepPunct{\mcitedefaultmidpunct}
{\mcitedefaultendpunct}{\mcitedefaultseppunct}\relax
\EndOfBibitem
\bibitem{Sjostrand:2006za}
T.~Sj\"{o}strand, S.~Mrenna, and P.~Skands,
  \ifthenelse{\boolean{articletitles}}{\emph{{PYTHIA 6.4 physics and manual}},
  }{}\href{https://doi.org/10.1088/1126-6708/2006/05/026}{JHEP \textbf{05}
  (2006) 026}, \href{http://arxiv.org/abs/hep-ph/0603175}{{\normalfont\ttfamily
  arXiv:hep-ph/0603175}}\relax
\mciteBstWouldAddEndPuncttrue
\mciteSetBstMidEndSepPunct{\mcitedefaultmidpunct}
{\mcitedefaultendpunct}{\mcitedefaultseppunct}\relax
\EndOfBibitem
\bibitem{LHCb-PROC-2010-056}
I.~Belyaev {\em et~al.}, \ifthenelse{\boolean{articletitles}}{\emph{{Handling
  of the generation of primary events in Gauss, the LHCb simulation
  framework}}, }{}\href{https://doi.org/10.1088/1742-6596/331/3/032047}{J.\
  Phys.\ Conf.\ Ser.\  \textbf{331} (2011) 032047}\relax
\mciteBstWouldAddEndPuncttrue
\mciteSetBstMidEndSepPunct{\mcitedefaultmidpunct}
{\mcitedefaultendpunct}{\mcitedefaultseppunct}\relax
\EndOfBibitem
\bibitem{Lange:2001uf}
D.~J. Lange, \ifthenelse{\boolean{articletitles}}{\emph{{The EvtGen particle
  decay simulation package}},
  }{}\href{https://doi.org/10.1016/S0168-9002(01)00089-4}{Nucl.\ Instrum.\
  Meth.\  \textbf{A462} (2001) 152}\relax
\mciteBstWouldAddEndPuncttrue
\mciteSetBstMidEndSepPunct{\mcitedefaultmidpunct}
{\mcitedefaultendpunct}{\mcitedefaultseppunct}\relax
\EndOfBibitem
\bibitem{Golonka:2005pn}
P.~Golonka and Z.~Was, \ifthenelse{\boolean{articletitles}}{\emph{{PHOTOS Monte
  Carlo: A precision tool for QED corrections in $Z$ and $W$ decays}},
  }{}\href{https://doi.org/10.1140/epjc/s2005-02396-4}{Eur.\ Phys.\ J.\
  \textbf{C45} (2006) 97},
  \href{http://arxiv.org/abs/hep-ph/0506026}{{\normalfont\ttfamily
  arXiv:hep-ph/0506026}}\relax
\mciteBstWouldAddEndPuncttrue
\mciteSetBstMidEndSepPunct{\mcitedefaultmidpunct}
{\mcitedefaultendpunct}{\mcitedefaultseppunct}\relax
\EndOfBibitem
\bibitem{Allison:2006ve}
Geant4 collaboration, J.~Allison {\em et~al.},
  \ifthenelse{\boolean{articletitles}}{\emph{{Geant4 developments and
  applications}}, }{}\href{https://doi.org/10.1109/TNS.2006.869826}{IEEE
  Trans.\ Nucl.\ Sci.\  \textbf{53} (2006) 270}\relax
\mciteBstWouldAddEndPuncttrue
\mciteSetBstMidEndSepPunct{\mcitedefaultmidpunct}
{\mcitedefaultendpunct}{\mcitedefaultseppunct}\relax
\EndOfBibitem
\bibitem{Agostinelli:2002hh}
Geant4 collaboration, S.~Agostinelli {\em et~al.},
  \ifthenelse{\boolean{articletitles}}{\emph{{Geant4: A simulation toolkit}},
  }{}\href{https://doi.org/10.1016/S0168-9002(03)01368-8}{Nucl.\ Instrum.\
  Meth.\  \textbf{A506} (2003) 250}\relax
\mciteBstWouldAddEndPuncttrue
\mciteSetBstMidEndSepPunct{\mcitedefaultmidpunct}
{\mcitedefaultendpunct}{\mcitedefaultseppunct}\relax
\EndOfBibitem
\bibitem{LHCb-PROC-2011-006}
M.~Clemencic {\em et~al.}, \ifthenelse{\boolean{articletitles}}{\emph{{The
  \lhcb simulation application, Gauss: Design, evolution and experience}},
  }{}\href{https://doi.org/10.1088/1742-6596/331/3/032023}{J.\ Phys.\ Conf.\
  Ser.\  \textbf{331} (2011) 032023}\relax
\mciteBstWouldAddEndPuncttrue
\mciteSetBstMidEndSepPunct{\mcitedefaultmidpunct}
{\mcitedefaultendpunct}{\mcitedefaultseppunct}\relax
\EndOfBibitem
\bibitem{LHCb-PAPER-2013-058}
LHCb collaboration, R.~Aaij {\em et~al.},
  \ifthenelse{\boolean{articletitles}}{\emph{{Study of forward $\Z$+jet
  production in $\proton\proton$ collisions at $\sqs = $7\tev}},
  }{}\href{https://doi.org/10.1007/JHEP01(2014)033}{JHEP \textbf{01} (2014)
  033}, \href{http://arxiv.org/abs/1310.8197}{{\normalfont\ttfamily
  arXiv:1310.8197}}\relax
\mciteBstWouldAddEndPuncttrue
\mciteSetBstMidEndSepPunct{\mcitedefaultmidpunct}
{\mcitedefaultendpunct}{\mcitedefaultseppunct}\relax
\EndOfBibitem
\bibitem{antikt}
M.~Cacciari, G.~P. Salam, and G.~Soyez,
  \ifthenelse{\boolean{articletitles}}{\emph{{The anti-kt jet clustering
  algorithm}}, }{}\href{https://doi.org/10.1088/1126-6708/2008/04/063}{JHEP
  \textbf{0804:063} (2008) },
  \href{http://arxiv.org/abs/0802.1189v}{{\normalfont\ttfamily
  arXiv:0802.1189v}}\relax
\mciteBstWouldAddEndPuncttrue
\mciteSetBstMidEndSepPunct{\mcitedefaultmidpunct}
{\mcitedefaultendpunct}{\mcitedefaultseppunct}\relax
\EndOfBibitem
\bibitem{fastjet}
M.~Cacciari, G.~P. Salam, and G.~Soyez,
  \ifthenelse{\boolean{articletitles}}{\emph{{FastJet user manual}},
  }{}\href{https://doi.org/10.1140/epjc/s10052-012-1896-2}{Eur.\ Phys.\ J.\
  \textbf{C72} (2012) 1896},
  \href{http://arxiv.org/abs/1111.6097}{{\normalfont\ttfamily
  arXiv:1111.6097}}\relax
\mciteBstWouldAddEndPuncttrue
\mciteSetBstMidEndSepPunct{\mcitedefaultmidpunct}
{\mcitedefaultendpunct}{\mcitedefaultseppunct}\relax
\EndOfBibitem
\bibitem{LHCb-PAPER-2015-016}
LHCb collaboration, R.~Aaij {\em et~al.},
  \ifthenelse{\boolean{articletitles}}{\emph{{Identification of beauty and
  charm quark jets at LHCb}},
  }{}\href{https://doi.org/10.1088/1748-0221/10/06/P06013}{JINST \textbf{10}
  (2015) P06013}, \href{http://arxiv.org/abs/1504.07670}{{\normalfont\ttfamily
  arXiv:1504.07670}}\relax
\mciteBstWouldAddEndPuncttrue
\mciteSetBstMidEndSepPunct{\mcitedefaultmidpunct}
{\mcitedefaultendpunct}{\mcitedefaultseppunct}\relax
\EndOfBibitem
\bibitem{Breiman}
L.~Breiman, J.~H. Friedman, R.~A. Olshen, and C.~J. Stone, {\em Classification
  and regression trees}, Wadsworth international group, Belmont, California,
  USA, 1984\relax
\mciteBstWouldAddEndPuncttrue
\mciteSetBstMidEndSepPunct{\mcitedefaultmidpunct}
{\mcitedefaultendpunct}{\mcitedefaultseppunct}\relax
\EndOfBibitem
\bibitem{Roe}
B.~P. Roe {\em et~al.}, \ifthenelse{\boolean{articletitles}}{\emph{{Boosted
  decision trees, an alternative to artificial neural networks}},
  }{}\href{https://doi.org/10.1016/j.nima.2004.12.018}{Nucl.\ Instrum.\ Meth.\
  \textbf{A543} (2005) 577},
  \href{http://arxiv.org/abs/physics/0408124}{{\normalfont\ttfamily
  arXiv:physics/0408124}}\relax
\mciteBstWouldAddEndPuncttrue
\mciteSetBstMidEndSepPunct{\mcitedefaultmidpunct}
{\mcitedefaultendpunct}{\mcitedefaultseppunct}\relax
\EndOfBibitem
\bibitem{AdaBoost}
Y.~Freund and R.~E. Schapire, \ifthenelse{\boolean{articletitles}}{\emph{A
  decision-theoretic generalization of on-line learning and an application to
  boosting}, }{}\href{https://doi.org/10.1006/jcss.1997.1504}{J.\ Comput.\
  Syst.\ Sci.\  \textbf{55} (1997) 119}\relax
\mciteBstWouldAddEndPuncttrue
\mciteSetBstMidEndSepPunct{\mcitedefaultmidpunct}
{\mcitedefaultendpunct}{\mcitedefaultseppunct}\relax
\EndOfBibitem
\bibitem{LHCB-PAPER-2016-011}
LHCb collaboration, R.~Aaij {\em et~al.},
  \ifthenelse{\boolean{articletitles}}{\emph{{Measurement of forward $\W\!$ and
  $\Z$ boson production in association with jets in proton-proton collisions at
  $\sqs=$8\tev}}, }{}\href{https://doi.org/10.1007/JHEP05(2016)131}{JHEP
  \textbf{05} (2016) 131},
  \href{http://arxiv.org/abs/1605.00951}{{\normalfont\ttfamily
  arXiv:1605.00951}}\relax
\mciteBstWouldAddEndPuncttrue
\mciteSetBstMidEndSepPunct{\mcitedefaultmidpunct}
{\mcitedefaultendpunct}{\mcitedefaultseppunct}\relax
\EndOfBibitem
\bibitem{Tikhonov:1963}
A.~N. Tikhonov, \ifthenelse{\boolean{articletitles}}{\emph{Solution of
  incorrectly formulated problems and the regularization method}, }{}Soviet
  Math.\ Dokl.\  \textbf{4} (1963) 1035\relax
\mciteBstWouldAddEndPuncttrue
\mciteSetBstMidEndSepPunct{\mcitedefaultmidpunct}
{\mcitedefaultendpunct}{\mcitedefaultseppunct}\relax
\EndOfBibitem
\bibitem{top}
LHCb collaboration, R.~Aaij {\em et~al.},
  \ifthenelse{\boolean{articletitles}}{\emph{{Measurement of forward top pair
  production in the dilepton channel in $pp$ collisions at $\sqs=$13\tev}},
  }{}\href{https://doi.org/10.1007/JHEP08(2018)174}{JHEP \textbf{08} (2018)
  174}, \href{http://arxiv.org/abs/1803.05188}{{\normalfont\ttfamily
  arXiv:1803.05188}}\relax
\mciteBstWouldAddEndPuncttrue
\mciteSetBstMidEndSepPunct{\mcitedefaultmidpunct}
{\mcitedefaultendpunct}{\mcitedefaultseppunct}\relax
\EndOfBibitem
\bibitem{bootstrap}
B.~Efron, \ifthenelse{\boolean{articletitles}}{\emph{{Bootstrap methods:
  Another look at jackknife}},
  }{}\href{https://doi.org/doi:10.1214/aos/1176344552}{The Annals of Statistics
  \textbf{7} (1979) 1}\relax
\mciteBstWouldAddEndPuncttrue
\mciteSetBstMidEndSepPunct{\mcitedefaultmidpunct}
{\mcitedefaultendpunct}{\mcitedefaultseppunct}\relax
\EndOfBibitem
\bibitem{mcstat}
I.~Narsky and F.~Porter,
  \ifthenelse{\boolean{articletitles}}{\emph{{Statistical analysis techniques
  in particle physics}},
  }{}\href{https://doi.org/10.1002/9783527677320}{Wiley-VCH (2013) }\relax
\mciteBstWouldAddEndPuncttrue
\mciteSetBstMidEndSepPunct{\mcitedefaultmidpunct}
{\mcitedefaultendpunct}{\mcitedefaultseppunct}\relax
\EndOfBibitem
\bibitem{tunfold}
S.~Schmitt, \ifthenelse{\boolean{articletitles}}{\emph{{TUnfold: an algorithm
  for correcting migration effects in high energy physics}},
  }{}\href{https://doi.org/10.1088/1748-0221/7/10/T10003}{JINST \textbf{7}
  (2012) T10003}, \href{http://arxiv.org/abs/1205.6201}{{\normalfont\ttfamily
  arXiv:1205.6201}}\relax
\mciteBstWouldAddEndPuncttrue
\mciteSetBstMidEndSepPunct{\mcitedefaultmidpunct}
{\mcitedefaultendpunct}{\mcitedefaultseppunct}\relax
\EndOfBibitem
\bibitem{lumi}
LHCb collaboration, R.~Aaij {\em et~al.},
  \ifthenelse{\boolean{articletitles}}{\emph{{Precision luminosity measurements
  at LHCb}}, }{}\href{https://doi.org/10.1088/1748-0221/9/12/P12005}{JINST
  \textbf{9} (2014) P12005},
  \href{http://arxiv.org/abs/1410.0149}{{\normalfont\ttfamily
  arXiv:1410.0149}}\relax
\mciteBstWouldAddEndPuncttrue
\mciteSetBstMidEndSepPunct{\mcitedefaultmidpunct}
{\mcitedefaultendpunct}{\mcitedefaultseppunct}\relax
\EndOfBibitem
\bibitem{nnpdf23}
NNPDF collaboration, R.~Abdul~Khalek {\em et~al.},
  \ifthenelse{\boolean{articletitles}}{\emph{{Parton distributions with theory
  uncertainties: General formalism and first phenomenological studies}},
  }{}\href{https://doi.org/10.1140/epjc/s10052-019-7401-4}{Eur.\ Phys.\ J.\
  \textbf{C79} (2019) 931},
  \href{http://arxiv.org/abs/1906.10698}{{\normalfont\ttfamily
  arXiv:1906.10698}}\relax
\mciteBstWouldAddEndPuncttrue
\mciteSetBstMidEndSepPunct{\mcitedefaultmidpunct}
{\mcitedefaultendpunct}{\mcitedefaultseppunct}\relax
\EndOfBibitem
\end{mcitethebibliography}
 
\newpage

\appendix
\appendixpage
\addappheadtotoc

\section{Differential cross sections as a function of leading jet $p_{\mathrm{T}}$ and $m_{jj}$ in linear scale}
\label{app:linear}

\begin{figure}[htb]
  \centering
    \includegraphics[width=0.49\linewidth]{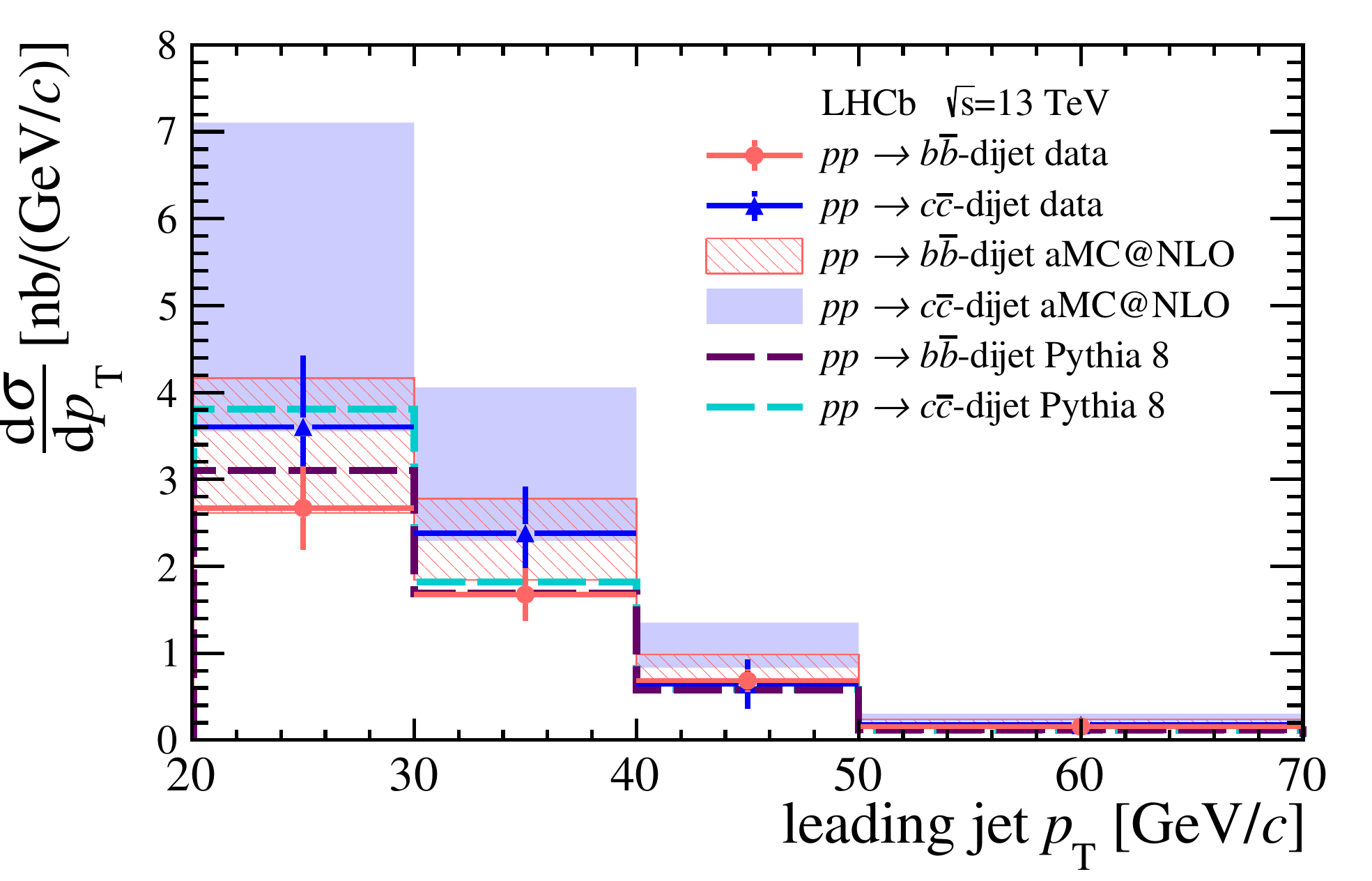} 
    \includegraphics[width=0.49\linewidth]{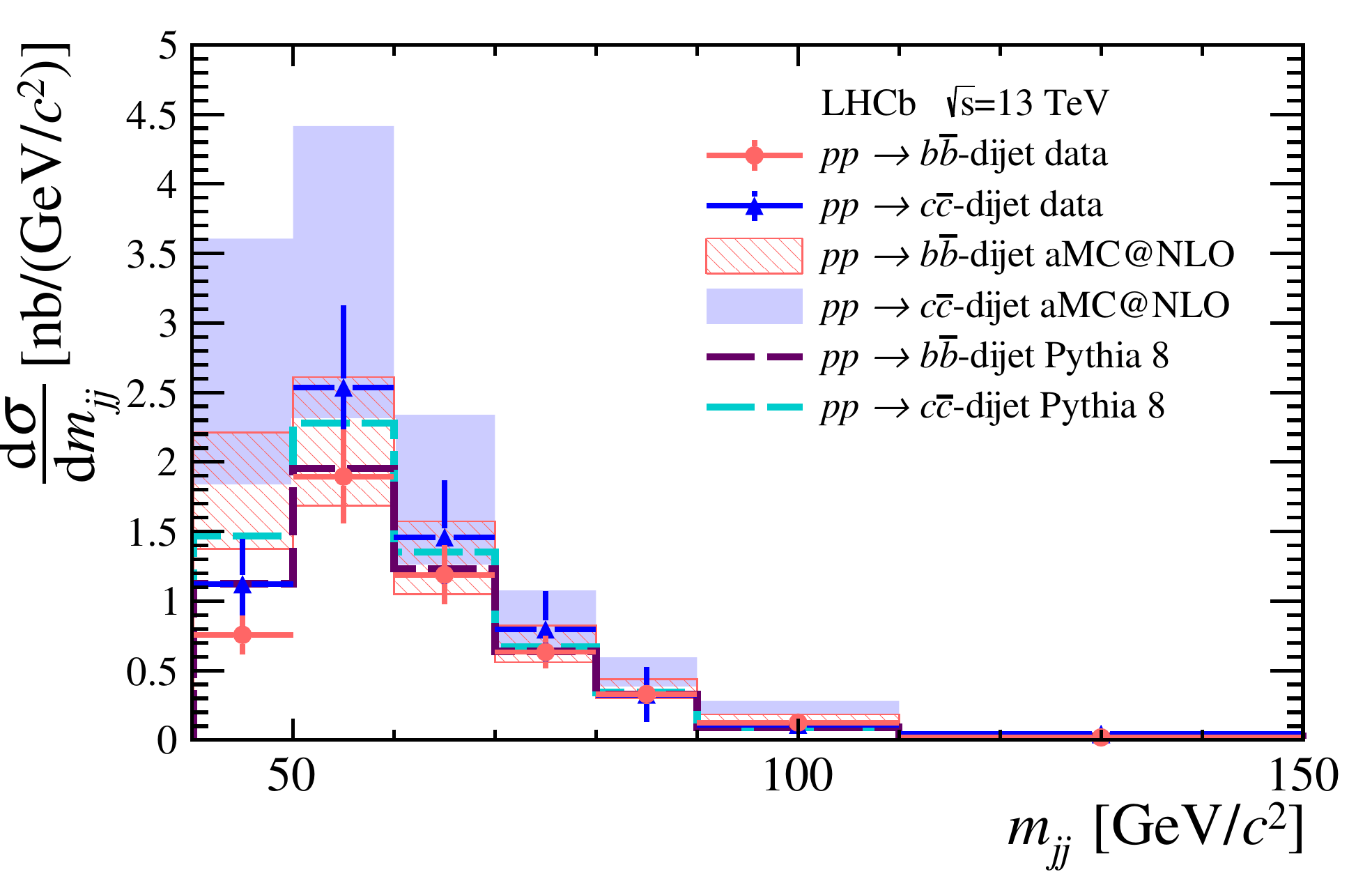}
  \caption{ Measured differential $b\bar{b}$- and $c\bar{c}$-dihet cross-sections as a function of the (left) leading jet $p_{\mathrm{T}}$ and (right) $m_{jj}$ on a linear scale. The error bars represent the total uncertainties, that are almost fully correlated across the bins. The next-to-leading-order predictions obtained with Madgraph5 aMC@NLO + \pythia are shown. The prediction uncertainty is dominated by the renormalisation and factorisation scale uncertainty. The leading-order prediction obtained with \pythia is also shown.
    \label{fig:xs_linear}}
\end{figure}

Figure \ref{fig:xs_linear} shows the $b\bar{b}$- and $c \bar{c}$-dijet differential cross-sections as a function of the leading jet $p_{\mathrm{T}}$ and $m_{jj}$ in linear scale.

\clearpage

\section{Numerical results and covariance matrices}
\label{app:cov}

The numerical values of the measured differential $b\bar{b}$- and $c \bar{c}$-dijet cross-sections, $c\bar{c}/b\bar{b}$ dijet cross-section ratios and their uncertainties are reported in Tables~\ref{tab:eta_results},~\ref{tab:ystar_results},~\ref{tab:pt_results} and \ref{tab:mass_results}. The covariance matrices for the $b\bar{b}$ ($c\bar{c}$) intervals and the cross-correlation matrix between $b\bar{b}$ and $c \bar{c}$ intervals are reported in Tables~\ref{tab:eta_bb_cov}--\ref{tab:mass_bb_cc_cov}. 

\begin{table}[h]
  \begin{center}
      \caption{\label{tab:eta_results} Numerical results of $b\bar{b}$- and $c \bar{c}$-dijet cross-sections, $c\bar{c}/b\bar{b}$ dijet cross-section ratios and their total uncertainties as a function of the leading jet $\eta$.}
      \begin{tabular}{c|c|c|c}
      \toprule
      $\eta$ bin & $\frac{\text{d}\sigma}{\text{d}\eta}(b\bar{b})$ [nb] & $\frac{\text{d}\sigma}{\text{d}\eta}(c\bar{c})$ [nb] & $R$ \\
      \midrule
 $[2.2,2.4]$ & 35.8 $\pm$ 7.0 & 62 $\pm$ 23 & 1.72 $\pm$ 0.70 \\
 $[2.4,2.6]$ & 39.9 $\pm$ 7.2 & 56 $\pm$ 17 & 1.41 $\pm$ 0.48 \\
 $[2.6,2.8]$ & 38.4 $\pm$ 6.8 & 52 $\pm$ 14 & 1.36 $\pm$ 0.41 \\
 $[2.8,3.0]$ & 35.1 $\pm$ 6.3 & 43 $\pm$ 10 & 1.24 $\pm$ 0.34 \\
 $[3.0,3.2]$ & 30.2 $\pm$ 5.4 & 39 $\pm$ \phantom{0}8 & 1.28 $\pm$ 0.34 \\
 $[3.2,3.4]$ & 25.1 $\pm$ 4.5 & 35 $\pm$ \phantom{0}8 & 1.40 $\pm$ 0.40 \\
 $[3.4,3.6]$ & 20.8 $\pm$ 3.8 & 26 $\pm$ \phantom{0}6 & 1.27 $\pm$ 0.36 \\
 $[3.6,3.8]$ & 16.6 $\pm$ 3.1 & 20 $\pm$ \phantom{0}5 & 1.21 $\pm$ 0.37 \\
 $[3.8,4.0]$ & 13.1 $\pm$ 2.5 & 14 $\pm$ \phantom{0}5 & 1.09 $\pm$ 0.42 \\
 $[4.0,4.2]$ & 10.5 $\pm$ 2.0 & 12 $\pm$ \phantom{0}7 & 1.17 $\pm$ 0.70 \\
        \bottomrule
      \end{tabular}
  \end{center}
\end{table}

\begin{table}[h]
  \begin{center}
      \caption{\label{tab:ystar_results} Numerical results of $b\bar{b}$- and $c \bar{c}$-dijet cross-sections, $c\bar{c}/b\bar{b}$ dijet cross-section ratios and their total uncertainties as a function of $\Delta y^*$.}
      \begin{tabular}{c|c|c|c}
      \toprule
      $\Delta y^*$ bin & $\frac{\text{d}\sigma}{\text{d}\Delta y^*}(b\bar{b})$ [nb] & $\frac{\text{d}\sigma}{\text{d}\Delta y^*}(c\bar{c})$ [nb] & $R$ \\
      \midrule
         $[0.0,0.1]$ & 104.5 $\pm$ 18.6 & 150.2 $\pm$ 34.0 & 1.4 $\pm$ 0.4 \\
 $[0.1,0.2]$ & \phantom{0}96.1 $\pm$ 17.1 & 131.8 $\pm$ 28.6 & 1.4 $\pm$ 0.4 \\
 $[0.2,0.3]$ & \phantom{0}85.7 $\pm$ 15.3 & 114.7 $\pm$ 24.7 & 1.3 $\pm$ 0.4 \\
 $[0.3,0.4]$ & \phantom{0}72.8 $\pm$ 13.0 & \phantom{0}96.1 $\pm$ 20.6 & 1.3 $\pm$ 0.4 \\
 $[0.4,0.5]$ & \phantom{0}59.7 $\pm$ 10.6 & \phantom{0}77.8 $\pm$ 16.9 & 1.3 $\pm$ 0.4 \\
 $[0.5,0.6]$ & \phantom{0}46.4 $\pm$ \phantom{0}8.3 &\phantom{0}62.9 $\pm$ 13.9 & 1.4 $\pm$ 0.4 \\
 $[0.6,0.7]$ & \phantom{0}32.3 $\pm$ \phantom{0}5.8 & \phantom{0}48.3 $\pm$ 11.0 & 1.5 $\pm$ 0.4 \\
 $[0.7,0.8]$ & \phantom{0}20.2 $\pm$ \phantom{0}3.7 & \phantom{0}27.4 $\pm$ \phantom{0}6.5 & 1.4 $\pm$ 0.4 \\
 $[0.8,0.9]$ & \phantom{00}9.3 $\pm$ \phantom{0}1.7 & \phantom{0}12.9 $\pm$ \phantom{0}3.4 & 1.4 $\pm$ 0.4 \\
 $[0.9,1.0]$ & \phantom{00}2.7 $\pm$ \phantom{0}0.5 & \phantom{00}3.9 $\pm$ \phantom{0}1.4 & 1.4 $\pm$ 0.6 \\
        \bottomrule
      \end{tabular}
  \end{center}
\end{table}

\begin{table}[h]
  \begin{center}
      \caption{\label{tab:pt_results} Numerical results of $b\bar{b}$- and $c \bar{c}$-dijet cross-sections, $c\bar{c}/b\bar{b}$ dijet cross-section ratios and their total uncertainties as a function of the leading jet $\pt$.}
      \begin{tabular}{c|c|c|c}
      \toprule
      $\pt$ bin [\gevc]& $\frac{\text{d}\sigma}{\text{d}\pt}(b\bar{b})$ [$\frac{\mathrm{nb}}{\gevc}$] & $\frac{\text{d}\sigma}{\text{d}\pt}(c\bar{c})$ [$\frac{\mathrm{nb}}{\gevc}$] & $R$ \\
      \midrule
 $[20,30]$ & 2.671 $\pm$ 0.481 & 3.60 $\pm$ 0.82 & 1.4 $\pm$ 0.4 \\
 $[30,40]$ & 1.685 $\pm$ 0.302 & 2.38 $\pm$ 0.54 & 1.4 $\pm$ 0.4 \\
 $[40,50]$ & 0.684 $\pm$ 0.130 & 0.64 $\pm$ 0.29 & 0.9 $\pm$ 0.5 \\
 $[50,70]$ & 0.155 $\pm$ 0.034 & 0.17 $\pm$ 0.08 & 1.1 $\pm$ 0.6 \\
        \bottomrule
      \end{tabular}
  \end{center}
\end{table}

\begin{table}[h]
  \begin{center}
      \caption{\label{tab:mass_results} Numerical results of $b\bar{b}$- and $c \bar{c}$-dijet cross-sections, $c\bar{c}/b\bar{b}$ dijet cross-section ratios and their total uncertainties as a function of $m_{jj}$.}
      \begin{tabular}{c|c|c|c}
      \toprule
      $m_{jj}$ bin [\gevcc]& $\frac{\text{d}\sigma}{\text{d}m_{jj}}(b\bar{b})$ [$\frac{\mathrm{nb}}{\gevcc}$] & $\frac{\text{d}\sigma}{\text{d}m_{jj}}(c\bar{c})$ [$\frac{\mathrm{nb}}{\gevcc}$] & $R$ \\
      \midrule
 $[40,50]$ & 0.757 $\pm$ 0.142 & 1.121 $\pm$ 0.327 & 1.5 $\pm$ 0.5 \\
 $[50,60]$ & 1.896 $\pm$ 0.340 & 2.536 $\pm$ 0.592 & 1.3 $\pm$ 0.4 \\
 $[60,70]$ & 1.189 $\pm$ 0.214 & 1.458 $\pm$ 0.411 & 1.2 $\pm$ 0.4 \\
 $[70,80]$ & 0.633 $\pm$ 0.119 & 0.796 $\pm$ 0.274 & 1.3 $\pm$ 0.5 \\
 $[80,90]$ & 0.331 $\pm$ 0.066 & 0.327 $\pm$ 0.198 & 1.0 $\pm$ 0.6 \\
 $[90,110]$ & 0.126 $\pm$ 0.026 & 0.112 $\pm$ 0.062 & 0.9 $\pm$ 0.5 \\
 $[110,150]$ & 0.019 $\pm$ 0.005 & 0.043 $\pm$ 0.021 & 2.2 $\pm$ 1.2 \\
        \bottomrule
      \end{tabular}
  \end{center}
\end{table}

\begin{landscape}
\begin{table}[h]
  \begin{center}
      \caption{\label{tab:eta_bb_cov} Covariance matrix, corresponding to the total uncertainties, obtained between the leading jet $\eta$ intervals of the $b\bar{b}$-dijet differential cross sections. The unit of all the elements of the matrix is nb$^2$.}
      \resizebox{20cm}{!}{
      \begin{tabular}{c|cccccccccc}
        \toprule
        & $[2.2,2.4]$ & $[2.4,2.6]$ & $[2.6,2.8]$ & $[2.8,3.0]$ & $[3.0,3.2]$ & $[3.2,3.4]$ & $[3.4,3.6]$ & $[3.6,3.8]$ & $[3.8,4.0]$ & $[4.0,4.2]$\\
\midrule
$[2.2,2.4]$ & 49.3 & 50.0 & 47.5 & 43.5 & 37.6 & 31.4 & 26.2 & 21.3 & 17.4 & 14.1 \\
$[2.4,2.6]$ & - & 51.7 & 48.6 & 44.5 & 38.5 & 32.2 & 26.8 & 21.8 & 17.8 & 14.4 \\
$[2.6,2.8]$ & - & - & 46.7 & 42.3 & 36.6 & 30.6 & 25.5 & 20.7 & 16.9 & 13.7 \\
$[2.8,3.0]$ & - & - & - & 39.1 & 33.5 & 28.0 & 23.3 & 18.9 & 15.5 & 12.5 \\
$[3.0,3.2]$ & - & - & - & - & 29.3 & 24.2 & 20.2 & 16.4 & 13.4 & 10.8 \\
$[3.2,3.4]$ & - & - & - & - & - & 20.4 & 16.9 & 13.7 & 11.2 & 9.0 \\
$[3.4,3.6]$ & - & - & - & - & - & - & 14.2 & 11.4 & 9.3 & 7.5 \\
$[3.6,3.8]$ & - & - & - & - & - & - & - & 9.4 & 7.6 & 6.1 \\
$[3.8,4.0]$ & - & - & - & - & - & - & - & - & 6.3 & 5.0 \\
$[4.0,4.2]$ & - & - & - & - & - & - & - & - & - & 4.1 \\
        \bottomrule
      \end{tabular}
      }
  \end{center}
\end{table}

\begin{table}[h]
  \begin{center}
      \caption{\label{tab:eta_cc_cov} Covariance matrix, corresponding to the total uncertainties, obtained between the leading jet $\eta$ intervals of the $c\bar{c}$-dijet differential cross sections. The unit of all the elements of the matrix is nb$^2$.}
      \resizebox{20cm}{!}{
      \begin{tabular}{c|cccccccccc}
        \toprule
       & $[2.2,2.4]$ & $[2.4,2.6]$ & $[2.6,2.8]$ & $[2.8,3.0]$ & $[3.0,3.2]$ & $[3.2,3.4]$ & $[3.4,3.6]$ & $[3.6,3.8]$ & $[3.8,4.0]$ & $[4.0,4.2]$\\
\midrule
$[2.2,2.4]$ & 519.3 & 388.0 & 306.7 & 218.8 & 190.1 & 186.7 & 143.2 & 117.8 & 110.4 & 157.3 \\
$[2.4,2.6]$ & - & 295.7 & 231.4 & 165.1 & 143.5 & 140.9 & 108.0 & 88.9 & 83.3 & 118.7 \\
$[2.6,2.8]$ & - & - & 184.8 & 130.5 & 113.4 & 111.3 & 85.4 & 70.3 & 65.9 & 93.8 \\
$[2.8,3.0]$ & - & - & - & 94.0 & 80.9 & 79.4 & 60.9 & 50.1 & 47.0 & 66.9 \\
$[3.0,3.2]$ & - & - & - & - & 71.0 & 69.0 & 53.0 & 43.6 & 40.8 & 58.2 \\
$[3.2,3.4]$ & - & - & - & - & - & 68.5 & 52.0 & 42.8 & 40.1 & 57.1 \\
$[3.4,3.6]$ & - & - & - & - & - & - & 40.3 & 32.8 & 30.8 & 43.8 \\
$[3.6,3.8]$ & - & - & - & - & - & - & - & 27.3 & 25.3 & 36.0 \\
$[3.8,4.0]$ & - & - & - & - & - & - & - & - & 24.0 & 33.8 \\
$[4.0,4.2]$ & - & - & - & - & - & - & - & - & - & 48.6 \\
        \bottomrule
      \end{tabular}
      }
  \end{center}
\end{table}

\begin{table}[h]
  \begin{center}
      \caption{\label{tab:eta_bb_cc_cov} Covariance matrix, corresponding to the total uncertainties, obtained between the leading jet $\eta$ intervals of the $b\bar{b}$ (horizontal) and $c\bar{c}$ (vertical) differential cross sections. The unit of all the elements of the matrix is nb$^2$.}
      \resizebox{20cm}{!}{
      \begin{tabular}{c|cccccccccc}
        \toprule
       & $[2.2,2.4]_{c\bar{c}}$ & $[2.4,2.6]_{c\bar{c}}$ & $[2.6,2.8]_{c\bar{c}}$ & $[2.8,3.0]_{c\bar{c}}$ & $[3.0,3.2]_{c\bar{c}}$ & $[3.2,3.4]_{c\bar{c}}$ & $[3.4,3.6]_{c\bar{c}}$ & $[3.6,3.8]_{c\bar{c}}$ & $[3.8,4.0]_{c\bar{c}}$ & $[4.0,4.2]_{c\bar{c}}$\\
       \midrule
$[2.2,2.4]_{b\bar{b}}$ & 0.63 & 0.48 & 0.38 & 0.27 & 0.23 & 0.23 & 0.18 & 0.15 & 0.14 & 0.19 \\
$[2.4,2.6]_{b\bar{b}}$ & 0.65 & 0.49 & 0.39 & 0.28 & 0.24 & 0.24 & 0.18 & 0.15 & 0.14 & 0.20 \\
$[2.6,2.8]_{b\bar{b}}$ & 0.62 & 0.47 & 0.37 & 0.26 & 0.23 & 0.22 & 0.17 & 0.14 & 0.13 & 0.19 \\
$[2.8,3.0]_{b\bar{b}}$ & 0.57 & 0.43 & 0.34 & 0.24 & 0.21 & 0.21 & 0.16 & 0.13 & 0.12 & 0.17 \\
$[3.0,3.2]_{b\bar{b}}$ & 0.49 & 0.37 & 0.29 & 0.21 & 0.18 & 0.18 & 0.14 & 0.11 & 0.11 & 0.15 \\
$[3.2,3.4]_{b\bar{b}}$ & 0.41 & 0.31 & 0.24 & 0.17 & 0.15 & 0.15 & 0.11 & 0.09 & 0.09 & 0.13 \\
$[3.4,3.6]_{b\bar{b}}$ & 0.34 & 0.26 & 0.20 & 0.15 & 0.13 & 0.12 & 0.09 & 0.08 & 0.07 & 0.10 \\
$[3.6,3.8]_{b\bar{b}}$ & 0.28 & 0.21 & 0.16 & 0.12 & 0.10 & 0.10 & 0.08 & 0.06 & 0.06 & 0.08 \\
$[3.8,4.0]_{b\bar{b}}$ & 0.23 & 0.17 & 0.13 & 0.10 & 0.08 & 0.08 & 0.06 & 0.05 & 0.05 & 0.07 \\
$[4.0,4.2]_{b\bar{b}}$ & 0.18 & 0.14 & 0.11 & 0.08 & 0.07 & 0.07 & 0.05 & 0.04 & 0.04 & 0.06 \\
        \bottomrule
      \end{tabular}
      }
  \end{center}
\end{table}

\begin{table}[h]
  \begin{center}
      \caption{\label{tab:ystar_bb_cov} Covariance matrix, corresponding to the total uncertainties, obtained between the $\Delta y^*$ intervals of the $b\bar{b}$-dijet differential cross sections. The unit of all the elements of the matrix is nb$^2$.}
      \resizebox{20cm}{!}{
      \begin{tabular}{c|cccccccccc}
        \toprule
      & $[0.0,0.1]$ & $[0.1,0.2]$ & $[0.2,0.3]$ & $[0.3,0.4]$ & $[0.4,0.5]$ & $[0.5,0.6]$ & $[0.6,0.7]$ & $[0.7,0.8]$ & $[0.8,0.9]$ & $[0.9,1.0]$\\
\midrule
$[0.0,0.1]$ & 346.36 & 315.22 & 281.25 & 239.06 & 196.04 & 152.71 & 106.49 & 67.22 & 31.46 & 9.87 \\
$[0.1,0.2]$ & - & 292.71 & 258.55 & 219.77 & 180.22 & 140.39 & 97.90 & 61.80 & 28.92 & 9.08 \\
$[0.2,0.3]$ & - & - & 233.02 & 196.08 & 160.80 & 125.26 & 87.35 & 55.14 & 25.81 & 8.10 \\
$[0.3,0.4]$ & - & - & - & 168.35 & 136.68 & 106.47 & 74.24 & 46.87 & 21.93 & 6.88 \\
$[0.4,0.5]$ & - & - & - & - & 113.22 & 87.31 & 60.88 & 38.43 & 17.99 & 5.64 \\
$[0.5,0.6]$ & - & - & - & - & - & 68.70 & 47.43 & 29.94 & 14.01 & 4.40 \\
$[0.6,0.7]$ & - & - & - & - & - & - & 33.41 & 20.88 & 9.77 & 3.07 \\
$[0.7,0.8]$ & - & - & - & - & - & - & - & 13.31 & 6.17 & 1.94 \\
$[0.8,0.9]$ & - & - & - & - & - & - & - & - & 2.92 & 0.91 \\
$[0.9,1.0]$ & - & - & - & - & - & - & - & - & - & 0.29 \\
        \bottomrule
      \end{tabular}
      }
  \end{center}
\end{table}

\begin{table}[h]
  \begin{center}
      \caption{\label{tab:ystar_cc_cov} Covariance matrix, corresponding to the total uncertainties, obtained between the $\Delta y^*$ intervals of the $c\bar{c}$-dijet differential cross sections. The unit of all the elements of the matrix is nb$^2$.}
      \resizebox{20cm}{!}{
      \begin{tabular}{c|cccccccccc}
        \toprule
        & $[0.0,0.1]$ & $[0.1,0.2]$ & $[0.2,0.3]$ & $[0.3,0.4]$ & $[0.4,0.5]$ & $[0.5,0.6]$ & $[0.6,0.7]$ & $[0.7,0.8]$ & $[0.8,0.9]$ & $[0.9,1.0]$\\
\midrule
$[0.0,0.1]$ & 1158.31 & 962.80 & 830.51 & 695.21 & 568.46 & 467.37 & 369.53 & 220.41 & 114.05 & 48.32 \\
$[0.1,0.2]$ & - & 816.54 & 697.31 & 583.71 & 477.28 & 392.41 & 310.26 & 185.06 & 95.75 & 40.57 \\
$[0.2,0.3]$ & - & - & 607.57 & 503.51 & 411.70 & 338.49 & 267.63 & 159.63 & 82.60 & 34.99 \\
$[0.3,0.4]$ & - & - & - & 425.74 & 344.63 & 283.35 & 224.03 & 133.63 & 69.14 & 29.29 \\
$[0.4,0.5]$ & - & - & - & - & 284.64 & 231.69 & 183.18 & 109.26 & 56.54 & 23.95 \\
$[0.5,0.6]$ & - & - & - & - & - & 192.41 & 150.61 & 89.83 & 46.48 & 19.69 \\
$[0.6,0.7]$ & - & - & - & - & - & - & 120.28 & 71.03 & 36.75 & 15.57 \\
$[0.7,0.8]$ & - & - & - & - & - & - & - & 42.79 & 21.92 & 9.29 \\
$[0.8,0.9]$ & - & - & - & - & - & - & - & - & 11.46 & 4.81 \\
$[0.9,1.0]$ & - & - & - & - & - & - & - & - & - & 2.06 \\
        \bottomrule
      \end{tabular}
      }
  \end{center}
\end{table}

\begin{table}[h]
  \begin{center}
      \caption{\label{tab:ystar_bb_cc_cov} Covariance matrix, corresponding to the total uncertainties, obtained between the $\Delta y^*$ intervals of the $b\bar{b}$ (horizontal) and $c\bar{c}$ (vertical)  differential cross sections. The unit of all the elements of the matrix is nb$^2$.}
      \resizebox{20cm}{!}{
      \begin{tabular}{c|cccccccccc}
        \toprule
        & $[0.0,0.1]_{c\bar{c}}$ & $[0.1,0.2]_{c\bar{c}}$ & $[0.2,0.3]_{c\bar{c}}$ & $[0.3,0.4]_{c\bar{c}}$ & $[0.4,0.5]_{c\bar{c}}$ & $[0.5,0.6]_{c\bar{c}}$ & $[0.6,0.7]_{c\bar{c}}$ & $[0.7,0.8]_{c\bar{c}}$ & $[0.8,0.9]_{c\bar{c}}$ & $[0.9,1.0]_{c\bar{c}}$\\
        \midrule
$[0.0,0.1]_{b\bar{b}}$ & 2.51 & 2.11 & 1.82 & 1.52 & 1.25 & 1.02 & 0.81 & 0.48 & 0.25 & 0.11 \\
$[0.1,0.2]_{b\bar{b}}$ & 2.31 & 1.94 & 1.67 & 1.40 & 1.15 & 0.94 & 0.74 & 0.44 & 0.23 & 0.10 \\
$[0.2,0.3]_{b\bar{b}}$ & 2.06 & 1.73 & 1.49 & 1.25 & 1.02 & 0.84 & 0.66 & 0.40 & 0.21 & 0.09 \\
$[0.3,0.4]_{b\bar{b}}$ & 1.75 & 1.47 & 1.27 & 1.06 & 0.87 & 0.71 & 0.56 & 0.34 & 0.17 & 0.07 \\
$[0.4,0.5]_{b\bar{b}}$ & 1.44 & 1.21 & 1.04 & 0.87 & 0.71 & 0.59 & 0.46 & 0.28 & 0.14 & 0.06 \\
$[0.5,0.6]_{b\bar{b}}$ & 1.12 & 0.94 & 0.81 & 0.68 & 0.56 & 0.46 & 0.36 & 0.22 & 0.11 & 0.05 \\
$[0.6,0.7]_{b\bar{b}}$ & 0.78 & 0.66 & 0.57 & 0.47 & 0.39 & 0.32 & 0.25 & 0.15 & 0.08 & 0.03 \\
$[0.7,0.8]_{b\bar{b}}$ & 0.49 & 0.41 & 0.36 & 0.30 & 0.24 & 0.20 & 0.16 & 0.09 & 0.05 & 0.02 \\
$[0.8,0.9]_{b\bar{b}}$ & 0.23 & 0.19 & 0.17 & 0.14 & 0.11 & 0.09 & 0.07 & 0.04 & 0.02 & 0.01 \\
$[0.9,1.0]_{b\bar{b}}$ & 0.07 & 0.06 & 0.05 & 0.04 & 0.04 & 0.03 & 0.02 & 0.01 & 0.01 & 0.00 \\

        \bottomrule
      \end{tabular}
      }
  \end{center}
\end{table}
\end{landscape}

\begin{table}[h]
  \begin{center}
      \caption{\label{tab:pt_bb_cov} Covariance matrix, corresponding to the total uncertainties, obtained between the leading jet $\pt$ intervals of the $b\bar{b}$-dijet differential cross sections. The unit of all the elements of the matrix is $\left( \frac{\mathrm{nb}}{\gevc} \right)^2$ and the $\pt$ intervals are given in \gevc.}
      \begin{tabular}{c|cccc}
      \toprule
        & $[20,30]$ & $[30,40]$ & $[40,50]$ & $[50,70]$\\
\midrule
$[20,30]$ & 0.2316 & 0.1440 & 0.0620 & 0.0162 \\
$[30,40]$ & - & 0.0913 & 0.0389 & 0.0102 \\
$[40,50]$ & - & - & 0.0169 & 0.0044 \\
$[50,70]$ & - & - & - & 0.0012 \\
        \bottomrule
      \end{tabular}
  \end{center}
\end{table}

\begin{table}[h]
  \begin{center}
      \caption{\label{tab:pt_cc_cov} Covariance matrix, corresponding to the total uncertainties, obtained between the leading jet $\pt$ intervals of the $c\bar{c}$-dijet differential cross sections. The unit of all the elements of the matrix is $\left( \frac{\mathrm{nb}}{\gevc} \right)^2$ and the $\pt$ intervals are given in \gevc.}
      \begin{tabular}{c|cccc}
        \toprule
         & $[20,30]$ & $[30,40]$ & $[40,50]$ & $[50,70]$\\
\midrule
$[20,30]$ & 0.6761 & 0.4386 & 0.2337 & 0.0665 \\
$[30,40]$ & - & 0.2903 & 0.1532 & 0.0436 \\
$[40,50]$ & - & - & 0.0824 & 0.0232 \\
$[50,70]$ & - & - & - & 0.0067 \\
        \bottomrule
      \end{tabular}
  \end{center}
\end{table}

\begin{table}[h]
  \begin{center}
      \caption{\label{tab:pt_bb_cc_cov} Covariance matrix, corresponding to the total uncertainties, obtained between the leading jet $\pt$ intervals of the $b\bar{b}$ (horizontal) and $c\bar{c}$ (vertical)  differential cross sections. The unit of all the elements of the matrix is $\left( \frac{\mathrm{nb}}{\gevc} \right)^2$ and the $\pt$ intervals are given in \gevc.}
      \begin{tabular}{c|cccc}
        \toprule
        & $[20,30]_{c\bar{c}}$ & $[30,40]_{c\bar{c}}$ & $[40,50]_{c\bar{c}}$ & $[50,70]_{c\bar{c}}$\\
        \midrule
$[20,30]_{b\bar{b}}$ & 0.0016 & 0.0010 & 0.0005 & 0.0002 \\
$[30,40]_{b\bar{b}}$ & 0.0010 & 0.0006 & 0.0003 & 0.0001 \\
$[40,50]_{b\bar{b}}$ & 0.0004 & 0.0003 & 0.0001 & 0.0000 \\
$[50,70]_{b\bar{b}}$ & 0.0001 & 0.0001 & 0.0000 & 0.0000 \\
        \bottomrule
      \end{tabular}
  \end{center}
\end{table}

\begin{landscape}
\begin{table}[h]
  \begin{center}
      \caption{\label{tab:mass_bb_cov} Covariance matrix, corresponding to the total uncertainties, obtained between the $m_{jj}$ intervals of the $b\bar{b}$-dijet differential cross sections. The unit of all the elements of the matrix is $\left( \frac{\mathrm{nb}}{\gev} \right)^2$ and the mass intervals are given in \gev.}
      \resizebox{14cm}{!}{
      \begin{tabular}{c|ccccccc}
        \toprule
        & $[40,50]$ & $[50,60]$ & $[60,70]$ & $[70,80]$ & $[80,90]$ & $[90,110]$ & $[110,150]$\\
\midrule
$[40,50]$ & 0.02005 & 0.04762 & 0.03002 & 0.01662 & 0.00928 & 0.00366 & 0.00077 \\
$[50,60]$ & - & 0.11536 & 0.07200 & 0.03987 & 0.02225 & 0.00878 & 0.00184 \\
$[60,70]$ & - & - & 0.04585 & 0.02513 & 0.01403 & 0.00553 & 0.00116 \\
$[70,80]$ & - & - & - & 0.01406 & 0.00777 & 0.00306 & 0.00064 \\
$[80,90]$ & - & - & - & - & 0.00438 & 0.00171 & 0.00036 \\
$[90,110]$ & - & - & - & - & - & 0.00068 & 0.00014 \\
$[110,150]$ & - & - & - & - & - & - & 0.00003 \\
        \bottomrule
      \end{tabular}
      }
  \end{center}
\end{table}

\begin{table}[h]
  \begin{center}
      \caption{\label{tab:mass_cc_cov} Covariance matrix, corresponding to the total uncertainties, obtained between the $m_{jj}$ intervals of the $c\bar{c}$-dijet differential cross sections. The unit of all the elements of the matrix is $\left( \frac{\mathrm{nb}}{\gevcc} \right)^2$ and the mass intervals are given in \gevcc.}
      \resizebox{14cm}{!}{
      \begin{tabular}{c|ccccccc}
        \toprule
        & $[40,50]$ & $[50,60]$ & $[60,70]$ & $[70,80]$ & $[80,90]$ & $[90,110]$ & $[110,150]$\\
\midrule
$[40,50]$ & 0.10704 & 0.19180 & 0.13302 & 0.08871 & 0.06419 & 0.02014 & 0.00694 \\
$[50,60]$ & - & 0.35066 & 0.24076 & 0.16056 & 0.11618 & 0.03646 & 0.01255 \\
$[60,70]$ & - & - & 0.16866 & 0.11135 & 0.08057 & 0.02528 & 0.00871 \\
$[70,80]$ & - & - & - & 0.07501 & 0.05373 & 0.01686 & 0.00581 \\
$[80,90]$ & - & - & - & - & 0.03927 & 0.01220 & 0.00420 \\
$[90,110]$ & - & - & - & - & - & 0.00387 & 0.00132 \\
$[110,150]$ & - & - & - & - & - & - & 0.00046 \\
        \bottomrule
      \end{tabular}
      }
  \end{center}
\end{table}

\begin{table}[h]
  \begin{center}
      \caption{\label{tab:mass_bb_cc_cov} Covariance matrix, corresponding to the total uncertainties, obtained between the $m_{jj}$ intervals of the $b\bar{b}$ (horizontal) and $c\bar{c}$ (vertical)  differential cross sections. The unit of all the elements of the matrix is $\left( \frac{\mathrm{nb}}{\gevcc} \right)^2$ and the mass intervals are given in \gevcc.}
      \resizebox{14cm}{!}{
      \begin{tabular}{c|ccccccc}
        \toprule
        & $[40,50]_{c\bar{c}}$ & $[50,60]_{c\bar{c}}$ & $[60,70]_{c\bar{c}}$ & $[70,80]_{c\bar{c}}$ & $[80,90]_{c\bar{c}}$ & $[90,100]_{c\bar{c}}$ & $[100,110]_{c\bar{c}}$\\
        \midrule
$[40,50]_{b\bar{b}}$ & 0.00018 & 0.00033 & 0.00023 & 0.00015 & 0.00011 & 0.00003 & 0.00001 \\
$[50,60]_{b\bar{b}}$ & 0.00044 & 0.00080 & 0.00055 & 0.00037 & 0.00027 & 0.00008 & 0.00003 \\
$[60,70]_{b\bar{b}}$ & 0.00028 & 0.00050 & 0.00035 & 0.00023 & 0.00017 & 0.00005 & 0.00002 \\
$[70,80]_{b\bar{b}}$ & 0.00015 & 0.00028 & 0.00019 & 0.00013 & 0.00009 & 0.00003 & 0.00001 \\
$[80,90]_{b\bar{b}}$ & 0.00009 & 0.00016 & 0.00011 & 0.00007 & 0.00005 & 0.00002 & 0.00001 \\
$[90,100]_{b\bar{b}}$ & 0.00003 & 0.00006 & 0.00004 & 0.00003 & 0.00002 & 0.00001 & 0.00000 \\
$[100,110]_{b\bar{b}}$ & 0.00001 & 0.00001 & 0.00001 & 0.00001 & 0.00000 & 0.00000 & 0.00000 \\
        \bottomrule
      \end{tabular}
      }
  \end{center}
\end{table}

\end{landscape}

\clearpage
\newpage
\centerline
{\large\bf LHCb collaboration}
\begin
{flushleft}
\small
R.~Aaij$^{31}$,
C.~Abell{\'a}n~Beteta$^{49}$,
T.~Ackernley$^{59}$,
B.~Adeva$^{45}$,
M.~Adinolfi$^{53}$,
H.~Afsharnia$^{9}$,
C.A.~Aidala$^{84}$,
S.~Aiola$^{25}$,
Z.~Ajaltouni$^{9}$,
S.~Akar$^{64}$
% ...

{\footnotesize \it

% ...
$ ^{9}$Universit{\'e} Clermont Auvergne, CNRS/IN2P3, LPC, Clermont-Ferrand, France\\
% ...
$ ^{25}$INFN Sezione di Milano, Milano, Italy\\
% ...
$ ^{31}$Nikhef National Institute for Subatomic Physics, Amsterdam, Netherlands\\
% ...
$ ^{45}$Instituto Galego de F{\'\i}sica de Altas Enerx{\'\i}as (IGFAE), Universidade de Santiago de Compostela, Santiago de Compostela, Spain\\
% ...
$ ^{49}$Physik-Institut, Universit{\"a}t Z{\"u}rich, Z{\"u}rich, Switzerland\\
% ...
$ ^{53}$H.H. Wills Physics Laboratory, University of Bristol, Bristol, United Kingdom\\
% ...
$ ^{59}$Oliver Lodge Laboratory, University of Liverpool, Liverpool, United Kingdom\\
% ...
$ ^{64}$University of Cincinnati, Cincinnati, OH, United States\\
% ...
$ ^{67}$Syracuse University, Syracuse, NY, United States\\
% ...
$ ^{84}$University of Michigan, Ann Arbor, United States, associated to $^{67}$\\
\bigskip
% ...
\medskip
}
\end{flushleft}

\end{document}